\documentclass[floats,floatfix,showpacs,preprintnumbers,amssymb,prd,superscriptaddress,nofootinbib,nolongbibliography,reprint]{revtex4-1}
\usepackage[dvipdfmx]{graphicx}

\usepackage{subfigure}
\usepackage{bm}
\usepackage{mathrsfs}
\usepackage{amsmath}
\usepackage{cases}

\def\be{\begin{equation}}
\def\ee{\end{equation}}
\newcommand{\beq}{\begin{eqnarray}}
\newcommand{\eeq}{\end{eqnarray}} 
 
\usepackage{color}
\usepackage[normalem]{ulem}
\usepackage[linktocpage=true]{hyperref}

\begin{document}
\title{On relativistic dynamical tides: subtleties and calibration}
\author{Takuya Katagiri}
\affiliation{Center of Gravity, Niels Bohr Institute, Blegdamsvej 17, 2100 Copenhagen, Denmark}
\author{Kent Yagi}
\affiliation{Department of Physics, University of Virginia, Charlottesville, Virginia 22904, USA}
\author{Vitor Cardoso}
\affiliation{Center of Gravity, Niels Bohr Institute, Blegdamsvej 17, 2100 Copenhagen, Denmark}

\affiliation{CENTRA, Departamento de F\'{\i}sica, Instituto Superior T\'ecnico -- IST, Universidade de Lisboa -- UL,
Avenida Rovisco Pais 1, 1049-001 Lisboa, Portugal}

\date{\today}

\begin{abstract}
The response of astrophysical compact objects to external tidal fields carries valuable information on the nature of these objects, on the equation of state of matter, and on the underlying gravitational theory.
In this work, we highlight subtleties in describing relativistic dynamical tidal responses that arise from ambiguities in the decomposition of a perturbed metric into external tidal and induced response pieces. Observables are unambiguous. However in practice, differences arising from implicit assumptions in the definition of tidal deformabilities may lead to a bias in constraining nuclear physics or gravitational theories, if not properly tied to observational data. We propose calibration of a tidal response function for any compact objects in vacuum General Relativity.  Within this framework, the dynamical tidal Love numbers of a Schwarzschild black hole in both even and odd sectors vanish at \emph{any} multipole order.  The calibration allows one to define dynamical tidal deformabilities of relativistic stars, such as neutron stars, as the difference from the BH values (zero) under the unified definition in a simple manner. As a straightforward extension of our framework, we compute the next-to-leading dissipative tidal response of Schwarzschild black holes for the first time. 
\end{abstract}

\maketitle


\section{Introduction}
\subsection{Gravitational-wave observations and tidal effects}

The detection of gravitational waves from a binary black hole (BH) merger opened up a new window of unprecedented opportunities to explore the Universe~\cite{LIGOScientific:2016aoc,LIGOScientific:2016vbw,Barack:2018yly}. Around $90$ events from the coalescence of binary BHs, binary BH--neutron stars, and binary neutron stars have been detected thus far~\cite{LIGOScientific:2018mvr,LIGOScientific:2020ibl,KAGRA:2021vkt}. The gravitational radiation emitted by compact binaries carries golden information concerning the geometry in the strong field, dynamical regime, allowing us to test the knowledge of gravity in an extreme environment~\cite{LIGOScientific:2016lio,Yunes:2016jcc,LIGOScientific:2018dkp,Barack:2018yly,Berti:2018cxi,Cardoso:2019rvt,Berti:2018vdi,LIGOScientific:2020tif}. All observations are consistent with theoretical predictions from General Relativity~(GR). Current and future detectors with increased sensitivity~\cite{LIGOScientific:2014pky,VIRGO:2014yos,Somiya:2011np,LISA:2017pwj,Punturo:2010zz,Reitze:2019iox} will provide exquisite data that may answer some of the outstanding questions in theoretical physics in the future~\cite{Barack:2018yly,Cardoso:2019rvt,Baibhav:2019rsa}.

Tidal effects in inspiralling binaries are promising phenomena leading to deep theoretical insights into the nature of compact objects~\cite{Cardoso:2017cfl,Kol:2011vg,Hui:2020xxx,Charalambous:2022rre,Hui:2021vcv,Katagiri:2022vyz}. To leading in the orbital angular velocity of the binary, the tidal deformation of a compact body, induced by the companion, is quantified by a set of {\it static tidal Love numbers}~(TLNs)~\cite{Hinderer:2007mb,Flanagan:2007ix,Binnington:2009bb,Damour:2009vw,poisson_will_2014}. The TLNs depend on the internal structure of a compact body as well as the underlying theory of gravity. In a post-Newtonian (PN) expansion of  gravitational waves from inspiralling binaries, their leading-order effect enters at 5PN order~\cite{Flanagan:2007ix}. Therefore, accurate modeling of tidal responses of compact objects is crucial for connecting gravitational-wave observations to theoretical astrophysics and nuclear physics. In particular, the measurement of TLNs allows us to test the multipolar structure of compact objects, say, neutron stars~\cite{Hinderer:2007mb,Yagi:2013bca,Yagi:2013sva}. In addition, TLNs are useful probes of exotic compact objects such as boson stars, gravastars,  or wormholes, which arise in different extensions of the standard model of particle physics or of GR~\cite{Cardoso:2017cfl}. Remarkably, BHs in vacuum GR have vanishing TLNs~\cite{Binnington:2009bb,Kol:2011vg,Cardoso:2017cfl,Cardoso:2018ptl,DeLuca:2022tkm,Hui:2020xxx,Poisson:2020vap,Poisson:2021yau,LeTiec:2020bos,LeTiec:2020spy,Chia:2020yla,Charalambous:2021mea}, while in the presence of matter fields or with deviation from GR, BHs can acquire nonzero TLNs~\cite{Cardoso:2017cfl,Cardoso:2019upw,Cardoso:2018ptl,DeLuca:2022xlz,Cardoso:2021wlq,DeLuca:2022tkm,Katagiri:2023yzm,Katagiri:2023umb}. Thus, the detection of nonzero TLNs could be a smoking gun for new physics in strong-field gravity.

Many of the studies of tidal physics in a general-relativistic setting focused on static (or adiabatic) tides, neglecting the time dependence of tides. The static-response approximation is appropriate during the earlier part of the late inspiral stage, when the orbital separation is not too small and changes in the internal structure occur on timescales much longer than the orbital period. However, as the separation shrinks due to gravitational-wave emission, the spacetime becomes highly dynamical. Then, the effect of {\it dynamical tides}, in which the time dependence of tidal responses during orbital motion is not negligible, may become significant for highly accurate and precise modeling of gravitational waveforms. Indeed, Refs.~\cite{Hinderer:2016eia,Steinhoff:2016rfi} pointed out the importance of including dynamical tides in waveform templates.

The dissipative part of the dynamical tidal effect involves {\it tidal heating}, also often referred to as tidal dissipation, which captures the exchange of energy and angular momentum of the body and an external tidal environment. The {\it tidal dissipation numbers}~(TDNs), which quantify tidal heating, first appear at 2.5PN~(4PN) order for a spinning~(nonspinning) body in a PN waveform~\cite{Poisson:1994yf,Tagoshi:1997jy,Alvi:2001mx,Poisson:2004cw}. For a BH with modest spin, the dissipative mechanism drives spin up, 
potentially giving rise to highly spinning BHs observable with gravitational-wave interferometers. If the BH spin is sufficiently high compared to a typical length scale of tidal environments, superradiant scattering is induced, leading to the extraction of BH energy and angular momentum during the inspiral stage. Tidal heating is important particularly in extreme-mass-ratio inspirals because the effect can contribute to thousands of orbital cycles if the primary is a BH~\cite{Hughes:2001jr,Price:2001un,Datta:2019euh,Datta:2019epe,Datta:2020rvo,Datta:2024vll}. Note also that the dissipative mechanism could be a discriminator of exotic compact objects with Planckian corrections at the horizon scale~\cite{Maselli:2017cmm}. The tidal heating allows one to infer the nature of the constituents of lower mass-gap binaries~\cite{Datta:2020gem}. It may be possible to precisely test the horizon presence or absence in third-generation detectors~\cite{Mukherjee:2022wws}.

The next-to-leading conservative 
tidal effect is quantified by a set of {\it dynamical tidal Love numbers}~(dTLNs), appearing first at 8PN order~\cite{Hinderer:2007mb,Hinderer:2016eia} in gravitational waveforms. Accurate waveform modeling at the late stage of the inspiral phase requires inclusion of the dynamical effects~\cite{Hinderer:2016eia,Steinhoff:2016rfi}. Neglecting dynamical tides may lead to a large systematic bias in determining tidal deformability and matter equation of state of neutron stars~\cite{Pratten:2021pro}. With dynamical tides, fundamental oscillation mode frequencies for neutron stars in GW170817 have been constrained~\cite{Pratten:2019sed}. Dynamical tides will further unveil important properties of compact objects~\cite{Poisson:2020vap,Katagiri:2023yzm,Chakraborty:2023zed,Saketh:2023bul,HegadeKR:2024agt}. 

Interest in dynamical tides is growing~\cite{Poisson:2020vap,HegadeKR:2024agt,Perry:2023wmm,Bhatt:2024yyz,Chakrabarti:2013lua,Saketh:2023bul,Chia:2024bwc}. However, as outlined below, there are subtleties in describing relativistic dynamical tidal effects, potentially leading to misinterpretation if one does not properly account for the differences between definitions used in computation of tidal responses and the tidal parameters incorporated in waveform models.

\subsection{Purpose of this work}
The aim of this work is precisely to highlight subtleties in the characterization of relativistic dynamical tidal effects and propose calibration of a tidal response function, allowing one to determine tidal responses for any compact objects in a unified and simple manner. In particular, we conduct a detailed study of the results derived in Ref.~\cite{Poisson:2020vap} and discuss ambiguities in the literature. We also discuss a physical interpretation of logarithmic running corrections to dTLNs for BHs reported in Refs.~\cite{Chakrabarti:2013lua,Saketh:2023bul,Perry:2023wmm,Chia:2024bwc}. For concreteness, we focus on non-spinning BHs in GR but the argument here is not restricted to spherical symmetry nor even to vacuum GR spacetimes~\cite{Katagiri:2024fpn}.

Subtleties in dTLNs arise from the ambiguity in the decomposition of a perturbed metric into external tidal and induced response pieces. Although the same issue occurs in static tides, the vanishing TLNs of Schwarzschild and Kerr BHs are derived with the use of a ``standard" definition based on the analytical properties of hypergeometric functions~\cite{Gralla:2017djj,Poisson:2020vap}. The ambiguity in dTLNs is caused by the absence of a ``standard" definition of relativistic dynamical tidal responses.
Additionally, literature often rely on the so-called near-zone approximation, which may fail to take some relevant terms into account, potentially giving a different result from a more robust analysis~\cite{Bhatt:2023zsy}.

In this work, we show that the dTLNs of a Schwarzschild BH in both even and odd sectors vanish at any multipole order within the proposed calibration, proving the conjecture by Poisson~\cite{Poisson:2020vap}. The calibration allows us to define dynamical tidal deformabilities of relativistic stars, such as neutron stars, as the difference from the BH zero values in a simple manner under the unified definition. We further present the next-to-leading TDNs for Schwarzschild BHs for the first time to the best of our knowledge.

Our study relies on a classical approach different from the recently developed worldline effective field theory~(EFT) framework~\cite{Saketh:2023bul,Ivanov:2024sds}, aiming to provide complementary insights. Our framework is designed to be compatible with PN framework~\cite{poisson_will_2014,Taylor:2008xy,Poisson:2018qqd,Poisson:2020vap}, which will offer an alternative pathway to connect tidal coefficients with observable quantities. While the EFT framework~\cite{Saketh:2023bul,Ivanov:2024sds} has already offered a certain level of understanding in this context, it is important to conduct equally detailed studies using the alternative method, bridging findings from both and reinforcing the foundation of relativistic tidal response theory.

\subsection{Organization of this paper}
The rest of this paper is organized as follows. In Sec.~\ref{Sec:TidallyDeformedbody}, we review the Newtonian and relativistic theory of tidal effects of spherically symmetric bodies. Section~\ref{Sec:MetricPerturbation} is devoted to providing the general solutions of the linearized Einstein equations in vacuum in terms of low-frequency expansions and then to explaining the matching procedure of a tidally perturbed Schwarzschild metric with a PN metric based on Refs.~\cite{poisson_will_2014,Taylor:2008xy,Poisson:2018qqd,Poisson:2020vap}. Calibration is introduced in Sec.~\ref{Sec:Calibration}. We analyze the dynamical tide of a Schwarzschild BH in Sec.~\ref{Sec:dynamicaltide} and then discuss the physical interpretation of the logarithmic running coefficients in Sec.~\ref{Sec:Running}. We summarize this work in Sec.~\ref{Sec:discussion}. Henceforth, we adopt geometric units~$c=G=1$ throughout the paper.

\section{Tidally deformed body}\label{Sec:TidallyDeformedbody}

\subsection{Newtonian case}\label{Sec:Newtonian}
In this section, we briefly review the description of tidal effects on a compact object~\cite{Taylor:2008xy,poisson_will_2014,Poisson:2018qqd,Poisson:2020vap}. We first focus on the Newtonian theory.
Consider a spherically symmetric object of mass $M$ and radius $r_0$, and place it in an externally sourced gravitational field, in vacuum. The mutual gravitational force is assumed to be weak, modeling an inspiralling binary system at sufficiently large separations. 

We work in the Fourier domain. In a Cartesian coordinate system~$x^i=(x,y,z)$ with the origin located at the center-of-mass of the body~$({\bm x}=0)$, the tidal environment around the body is characterized by the {\it tidal moments},
\begin{equation}
\label{eq:tidalmomentsinNewton}
{\cal E}_{L}\left(\omega\right):=\left[\partial_{L}U_{\rm ext}\left(\omega;{\bm x}\right)\big|_{{\bm x}=0}\right]^{\rm STF},
\end{equation}
where $U_{\rm ext}(\omega;{\bm x})$ is the gravitational potential created by the external source. Here, $L:=i_1 i_2\cdots i_\ell$ stands for a collection of $\ell$ individual indices; for any $\ell$-index tensor, STF means that the tensor is symmetric and tracefree in any pair of arbitrary two indices in $L$~\cite{poisson_will_2014}. The {\it multipole moments} of the body are defined as
\begin{equation}
I^{L}\left(\omega\right):=\left[\int \rho\left(\omega;{\bm x}'\right) x'^{L}d^3x'\right]^{\rm STF}\,,\label{multipolemoments}
\end{equation}
where $\rho(\omega;{\bm x})$ is the mass density inside the body. In the absence of the external tidal field, the body is spherically symmetric and all multipole moments with $\ell\neq0 $ vanish. 

We assume that the tidal response can be modeled within a linear perturbation theory. In the presence of the tidal field, higher multipole moments beyond the dipole order are induced, which are linearly related to the tidal moments~\eqref{eq:tidalmomentsinNewton}:
\begin{equation}
\label{eq:QLFEL}
I^{L}\left(\omega\right)=-\frac{2}{\left(2\ell-1\right)!!}r_0^{2\ell+1}\hat{\cal F}_{\ell}(\omega){\cal E}^L\left(\omega\right).
\end{equation}
Here $\hat{\cal F}_{\ell}(\omega)$
is a dimensionless function of $\omega$ called a {\it tidal response function} which quantifies the tidal response of the body. The description in an STF tensor is translated into a spherical-harmonic description with the relations~\cite{poisson_will_2014},
\begin{align}
    I^L\left(\omega\right)=&\frac{4\pi \ell!}{\left(2\ell+1\right)!!}\sum_{m=-\ell}^\ell \mathscr{ Y}_{\ell m}^{*\langle L\rangle}\hat{I}_{\ell m}\left(\omega\right),\\
    {\cal E}^L\left(\omega\right)=&-\frac{4\pi \ell!}{2\ell+1}\sum_{m=-\ell}^\ell \mathscr{Y}_{\ell m}^{*\langle L\rangle} \hat{d}_{\ell m}\left(\omega\right) ,
\end{align}
where $\mathscr{Y}_{\ell m}^{*\langle L\rangle}$ is the coefficient of the expansion of spherical harmonics in STF tensors in a Cartesian coordinate system. Equation~\eqref{eq:QLFEL} is then simplified to
\begin{equation}
\label{eq:QFd}
\hat{I}_{\ell m}\left(\omega\right)=2 r_0^{2\ell+1}\hat{\cal F}_{\ell}(\omega) \hat{d}_{\ell m}\left(\omega\right).
\end{equation}

The tidal response function~$\hat{\cal F}_{\ell}(\omega)$ is, in general, a complex function and depends on the internal structure of the body, orbital frequency, and the underlying theory of gravity. Its real and imaginary parts describe conservative and dissipative interactions during orbital motion, respectively. Assuming that $1/\omega$ is much larger than $M$, we expand $\hat{\cal F}_{\ell}(\omega)$ as
\begin{align}
    \hat{\cal F}_{\ell}(\omega)=\kappa_{\ell}^{(0)}+i\omega M\nu_{\ell}^{(0)}+\left(\omega M\right)^2 \kappa_{\ell}^{(1)}+{\cal O}(\omega^3)\,.\label{eq:TLNsTDNsdTLNs}
\end{align}
We here introduced {\it tidal Love numbers}~(TLNs) $\kappa_{\ell}^{(0)}$, {\it tidal dissipation numbers}~(TDNs) $\nu_{\ell}^{(0)}$ , and {\it dynamical tidal Love numbers}~(dTLNs)~$\kappa_{\ell}^{(1)}$.

\subsection{Relativistic case}\label{Sec:Relativistic}
Relativistic tidal effects can be studied with matched asymptotic expansions~\cite{Detweiler:2005kq,Taylor:2008xy,Poisson:2018qqd,Poisson:2020vap,HegadeKR:2024agt}. We continue with the assumption that the mutual gravitational interaction (between two compact bodies in a binary) is weak and caused by two compact bodies far from each other. We focus on one of these two bodies, of mass $M$. The assumption of large separation is translated into the condition, $M\ll {\cal R}$ for the characteristic length scale~${\cal R}$ of the external universe, allowing to separate the outer region of the body into two: a {\it body zone}~$(M \lesssim r\ll {\cal R})$, which is a neighborhood of the body with strong gravity; a {\it PN zone}~($M\ll r\ll {\cal R}$), where gravity is weak and is described by a PN expansion of the external universe metric.    

In the body zone, the object is described by full GR. The tidally deformed body obeys the linearized Einstein equations. The multipole structure is determined by solving the linearized Einstein equations but the tidal moments are not specified. The multipole moments for stationary and asymptotically flat spacetimes are defined from the spacetime structure at large distances in two independent but equivalent manners~\cite{Geroch:1970cd,Hansen:1974zz,RevModPhys.52.299,1983GReGr..15..737G}. In spite of the gauge-invariant definition in the above works, the identification of multipole moments from metric perturbations is ambiguous~\cite{Gralla:2017djj,Poisson:2020vap}. In particular, any subtraction of an ``external tidal field" from the perturbed metric (to obtain an asymptotically flat metric) requires a subjective and arbitrary definition of a tidal field. Therefore, we adopt the operational definition used by Refs.~\cite{Poisson:2020vap,Pitre:2023xsr,HegadeKR:2024agt}: the induced multipole moments are a property of the body from the point of view of a point mass moving in a PN spacetime. The matching stated below then determines the value of the induced multipole moments without the above artificial subtraction.

In the PN zone, a body is viewed as a point mass with a multipole structure, moving on a trajectory in an ambient weak field. In this case, the multipole moments are not specified, while the tidal moments are given. The latter are defined in terms of the frame components of the covariant derivatives of the Weyl tensor for a local metric expanded around the worldline of the point mass along a timelike geodesic~\cite{Thorne:1984mz,Zhang:1986cpa,Detweiler:2005kq,Poisson:2020vap,Poisson:2018qqd,Taylor:2008xy,Poisson:2009qj}:
\begin{align}
\label{eq:eletidalmoments}
    {\cal E}_{L}\left(t\right):=&\frac{1}{\left(\ell-2\right)!}\left[C_{0i_10 i_2|i_3\cdots i_\ell}\right]^{\rm STF},\\
    {\cal B}_{L}\left(t\right):=&\frac{3}{2\left(\ell+1\right)\left(\ell-2\right)!}\left[\varepsilon_{i_1jk} C^{jk}_{~~i_20|i_3\cdots i_\ell}\right]^{\rm STF},\label{eq:magtidalmoments}
\end{align}
where $t$ is a proper time; $\varepsilon$ is the permutation symbol; $|$ denotes covariant derivatives projected to the frame components along the geodesic. The local metric matches with the massless limit of a body-zone metric~\cite{Thorne:1984mz}.

Each zone is described in a different approach but both are a part of the same spacetime. Two descriptions, therefore, must agree in an overlapping zone, where both are valid, in the same coordinate system. The body zone extends up to the PN zone because of the presence of a {\it buffer zone}~($M\ll r\ll {\cal R}$) between the body zone and the PN zone. Likewise, the PN zone extends to the body zone in the buffer zone. Asymptotically matching the two metrics in each region in the same gauge and the same coordinates in the buffer zone determines unspecified constants in each region approximately, which are linearly related via Eq.~\eqref{eq:QFd}. That is, the PN information determines the tidal moments of the body in the body zone, and the body-zone information determines the multipole moments of the body in the PN zone~\cite{Taylor:2008xy,Poisson:2020vap,HegadeKR:2024agt}.
 
Note that, although Eqs.~\eqref{eq:eletidalmoments} and~\eqref{eq:magtidalmoments} are meant to represent tidal moments created by an external source, the interpretation is just an approximation within matched asymptotic expansions. This is because~(i) a body has a finite size and the worldline needs to be replaced by a worldtube;~(ii) a body does not follow a geodesic in general;~(iii)~the Weyl tensor of a body's metric includes the body's own contribution. A tidally deformed metric is a superposition of tidal field and response pieces; both are essential to describe the tidally perturbed spacetime accurately. There is no unambiguous way to decompose the perturbed metric into the external tidal and body's response pieces.

\section{Metric perturbation of a spherically symmetric spacetime}\label{Sec:MetricPerturbation}
\subsection{Linearized Einstein equations}
Here, we apply the above framework to a tidally deformed Schwarzschild geometry, and then match it with a PN metric, connecting integration constants to quantities characterizing the tidal response. Consider a Schwarzschild spacetime in static $(t,r,\theta,\varphi)$ coordinates,
\begin{equation}
g_{\mu\nu}^{(0)}dx^\mu dx^\nu=-fdt^2+f^{-1}dr^2+r^2d\Omega^2\,,\label{eq:Schwarzschildmetric}
\end{equation}
with $f=1-2M/r$ and the line element of a unit two-sphere $d\Omega^2:=d\theta^2+\sin^2\theta d\varphi^2$. Consider an even-parity linear metric perturbation,~$g_{\mu\nu}= g_{\mu\nu}^{(0)}+h_{\mu\nu}^{\rm (even)}$ (we discuss the odd-parity sector in Appendix~\ref{Appendix:LowFrequencyExpansion}). In the Regge-Wheeler gauge~\cite{PhysRev.108.1063}, $h_{\mu\nu}^{\rm (even)}$ with a spherical harmonic decomposition in Fourier domain reads
\begin{align}
    h_{\mu\nu}^{\rm (even)}dx^\mu dx^\nu= &
\bigg(
f H dt^2-2i H_1 dtdr \nonumber \\
&-f^{-1}H_2 dr^2-r^2 K d\Omega^2\bigg) Y_{\ell m}e^{-i\omega t}\,,\label{hmunueven}
\end{align}
where $H=H(r)$, $H_i=H_i(r)$, $K=K(r)$, and $Y_{\ell m}=Y_{\ell m}(\theta,\varphi)$ are scalar spherical harmonics. 

Linearizing Einstein equations in vacuum,~$\delta G_{\mu\nu}=0$, we obtain coupled differential equations. The equations governing the even-parity metric perturbation are summarized into
\begin{align}
\left(\frac{d^2}{dr^2}+W_1^+\frac{d}{dr}+ W_2^+ \right)H=&0.\label{eq:LinearEveneq}
\end{align}
Explicit forms of $W_1^+$ and $W_2^+$ are given in Appendix~\ref{Appendix:ExplicitForms}.

\subsection{Low-frequency expansion}

We solve Eq.~\eqref{eq:LinearEveneq} with a low-frequency expansion. In Appendix~\ref{Appendix:LowFrequencyExpansion}, we provide the technical details of the low-frequency expansion in both even and odd parities. Now, expand $H$ in $\omega M$ as ($H^{(j)}=H^{(j)}(r)$)
\begin{align}
H=&H^{(0)}+\omega M H^{(1)}+\left(\omega M\right)^2 H^{(2)} +\mathcal{O}((\omega M)^3)\,.
\end{align}
Then, Eq.~\eqref{eq:LinearEveneq} reduces to
\begin{align}
{\cal L}_\ell^+\left[H^{(0)} \right]=&0,\label{eq:eqforH0}\\
{\cal L}_\ell^+\left[H^{(1)}  \right]=&0,\label{eq:eqforH1}\\
{\cal L}_\ell^+\left[H^{(2)}  \right]=&{\cal S}_\ell^+\left[H^{(0)}\right]\label{eq:eqforH2},
\end{align}
with the derivative operator,
\begin{align}
{\cal L}_\ell^+:=&\frac{d^2}{dr^2}+\frac{2\left(r-M\right)}{r^2f}\frac{d}{dr}-\frac{1}{r^2f}\left[\ell\left(\ell+1\right)+\frac{4M^2}{r^2f}\right].
\end{align}
The explicit form of ${\cal S}_\ell^+$ is given in Eq.~\eqref{eq:Tp}. Notice that ${\cal S}_\ell^+\left[H^{(0)}\right]$ is linear in $H^{(0)}$. Analytical properties of $H^{(0)}$, $H^{(1)}$, and $H^{(2)}$ are provided in Appendix~\ref{Appendix:AnalyticalProperties}.

The analytic expression for $H$ up to second order in $\omega M$ takes the form,
\begin{align}
\label{eq:H0general}
H=&\left[\mathbb{E}_{\ell m}^{(0)}+\omega M \mathbb{E}_{\ell m}^{(1)}+\left(\omega M\right)^2 \mathbb{E}_{\ell m}^{(2)}\right]H_\ell^T\nonumber\\
&+\left[\mathbb{I}_{\ell m}^{+(0)}+\omega M \mathbb{I}_{\ell m}^{+(1)}+\left(\omega M\right)^2\mathbb{I}_{\ell m}^{+(2)}\right]H_\ell^R\\
&+\left(\omega M\right)^2 P_\ell^++{\cal O}\left(\left(\omega M\right)^3\right).\nonumber
\end{align}
Here, we have defined $H_\ell^T$, $H_\ell^R$, and $P_\ell^+$ as
\begin{align}
    H_\ell^T:=& f\left(\frac{r}{M}\right)^\ell \!~_2F_1\left(-\ell+2,-\ell;-2\ell;2M/r\right),\label{eq:HT}\\
   H_\ell^R:=&f\left(\frac{M}{r}\right)^{\ell+1}\!~_2F_1\left(\ell+1,\ell+3;2\ell+2;2M/r\right),\label{eq:HR}\\
     P_\ell^+:=&\mathbb{E}_{\ell m}^{(0)} P_\ell^{+T}+\mathbb{I}_{\ell m}^{+(0)} P_\ell^{+R},\label{eq:Peven}
\end{align}
where $\!_2F_1(a,b;c;2M/r)$ is a Gaussian hypergeometric function~\cite{NIST:DLMF}. The coefficients $\mathbb{E}_{\ell m}^{(j)}$, $\mathbb{I}_{\ell m}^{+(j)}$ are integration constants (of Eqs.~\eqref{eq:eqforH0}--\eqref{eq:eqforH2}). Simpler, closed-form expressions for $H_2^{T}$ and $H_2^R$ are given in Eq.~\eqref{eq:homogeneoussols}. The solution~\eqref{eq:H0general} is consistent with the general solution of the linearized Einstein equations in Refs.~\cite{Poisson:2020vap,HegadeKR:2024agt}. 

The functions~$P_\ell^{+T}$ and $P_\ell^{+R}$ in Eq.~\eqref{eq:Peven} arise from particular solutions of Eq.~\eqref{eq:eqforH2} and are given by 
\begin{align}
P_\ell^{+T}&= I_\ell^{+TT}  H_\ell^R-  I_\ell^{+RT} H_\ell^T,\label{eq:PT}\\
P_\ell^{+R}&= I_\ell^{+TR}   H_\ell^R- I_\ell^{+RR} H_\ell^T\,.\label{eq:PR}
\end{align}
Here, $I_\ell^{+ij}=I_\ell^{+ij}(r)$  with $i,j$ running $T,R$ satisfy
\begin{align}
    I_\ell^{+ij}(r)=\int \frac{H_\ell^i \left(r\right){\cal S}_\ell^+\left[H_\ell^j\right]}{{\cal W}^+}dr,~~  {\cal W}^+:=&-\frac{2\ell+1}{fM}\left(\frac{M}{r}\right)^2.
\end{align}
The explicit forms of $P_2^{+T}$ and $P_2^{+R}$ are given in Eqs.~\eqref{eq:PT2} and~\eqref{eq:PR2}, respectively. Let us emphasize that the functional form of the particular solutions is not uniquely specified because one can add/subtract homogeneous pieces,~$H_\ell^{T}$ and $H_\ell^{R}$, while the solutions still solve Eq.~\eqref{eq:eqforH2}, allowing the arbitrary shift of  $\mathbb{E}_{\ell m}^{(2)}$ and $\mathbb{I}_{\ell m}^{+(2)}$. Physically, their ambiguity is attributed to that in the decomposition of the metric perturbation into external tidal and induced response pieces.

\subsection{Boundary condition at large distances}\label{sec:PNmatching}
We impose a boundary condition at large distances on Eq.~\eqref{eq:H0general} in terms of matching with a PN metric at the leading order following Refs.~\cite{Poisson:2018qqd,Poisson:2020vap,HegadeKR:2024agt}. To do so, we now change from the coordinates~$(t,r,\theta,\varphi)$ into a harmonic coordinate system,~$(t,\bar{x}^a)$ in a Schwarzschild spacetime, defined by
\begin{align}
\label{eq:HarmonicCoordinate}
    \bar{x}^a=&\bar{r} \Omega^a,~~\bar{r}:=r-M,\\
    \Omega^a=&\left(\sin \theta \cos\varphi,\sin\theta \sin\varphi,\cos\theta\right).\nonumber
\end{align}
Note that this coordinate system is not exactly harmonic in a tidally perturbed spacetime but the violation should be small under the assumption of the weak and low-frequency tidal perturbation. In the coordinate system~\eqref{eq:HarmonicCoordinate}, the tidally deformed Schwarzschild metric admits a PN expansion compatible with a PN metric in $M\ll \bar{r}\ll 1/\omega $:
\begin{widetext}
\begin{align}
\label{eq:expansiongtt}
    g_{tt}\big|_{M\ll\bar{r}\ll1/\omega}=&-1+\frac{2M}{\bar{r}}+\left[\mathbb{E}_{\ell m}^{(0)}+\omega M \mathbb{E}_{\ell m}^{(1)}+\left(\omega M\right)^2\mathbb{E}_{\ell m}^{(2)}\right]\left(\frac{\bar{r}}{M}\right)^\ell Y_{\ell m}e^{-i \omega t}\nonumber \\
    &+\left[\mathbb{I}_{\ell m}^{+(0)}+\omega M \mathbb{I}_{\ell m}^{+(1)}+\left(\omega M\right)^2\mathbb{I}_{\ell m}^{+(2)}\right]\left(\frac{M}{\bar{r}}\right)^{\ell+1} Y_{\ell m}e^{-i \omega t}\\
    &+{\cal O}\left(1/c^4, \bar{r}^{\ell-1},\bar{r}^{-\ell-2},\omega^2 \bar{r}^{\ell+2},\omega^2 \bar{r}^{-\ell+1}, \omega^2 \bar{r}^{\ell}\ln\bar{r},\omega^2\bar{r}^{-\ell-1}\ln\bar{r}\right).\nonumber
\end{align}
\end{widetext}
Now, one observes the emergence of logarithmic terms of $\bar{r}$ at ${\cal O}(\omega^2M^2)$, which originate from $P_\ell^{+T}$ and $P_\ell^{+R}$~(see Eqs.~\eqref{eq:largePT} and~\eqref{eq:largePR} for instance). In Sec.~\ref{Sec:Running}, we will discuss the physical interpretation for the origin of these logarithmic terms.

In the PN zone, the body is viewed as a point mass with multipole moments. The metric of the external universe in the harmonic coordinates~$(t,\bar{x}^a)$ is expanded as~\cite{Taylor:2008xy,Poisson:2018qqd,Poisson:2020vap,HegadeKR:2024agt}
\begin{eqnarray}
g_{tt}&=&-1+2\bar{U}(t,\bar{x}^a)+{\cal O}\left(1/c^4\right),\nonumber\\
g_{ta}&=&{\cal O}\left(1/c^3\right)\,,\, g_{ab}= \delta_{ab}+{\cal O}\left(1/c^2\right),\label{eq:PNmetricinTX}
\end{eqnarray}
with the Newtonian gravitational potential of the system,
\begin{align}
\bar{U}\left(t,\bar{x}^a\right)=\frac{M}{\bar{r}}+\frac{4\pi Y_{\ell m}}{2\ell+1}\left(\hat{d}_{\ell m}\bar{r}^\ell+\frac{\hat{I}_{\ell m }}{\bar{r}^{\ell+1}}\right)e^{-i\omega t}.\label{eq:tbarxa}
\end{align}
Here, $\hat{d}_{\ell m}=\hat{d}_{\ell m}(\omega)$ and $\hat{I}_{\ell m}=\hat{I}_{\ell m}(\omega)$ are the tidal moment and the multipole moment of the object, respectively.

Matching Eqs.~\eqref{eq:expansiongtt} and~\eqref{eq:PNmetricinTX} yields
\begin{eqnarray}
    \mathbb{E}_{\ell m}^{(0)}+\omega M \mathbb{E}_{\ell m}^{(1)}+\left(\omega M\right)^2\mathbb{E}_{\ell m}^{(2)}=&\dfrac{8\pi M^\ell }{2\ell+1}\hat{d}_{\ell m}\,,\label{eq:TidalMoment}\\
    \mathbb{I}_{\ell m}^{+(0)}+\omega M \mathbb{I}_{\ell m}^{+(1)}+\left(\omega M\right)^2\mathbb{I}_{\ell m}^{+(2)}=&\dfrac{8\pi M^{-\ell-1}}{2\ell+1}\hat{I}_{\ell m}.\label{eq:matching}
\end{eqnarray}
We thus rewrite Eq.~\eqref{eq:H0general} into
\begin{align}
    H\left(r\right)=&\frac{8\pi M^\ell }{2\ell+1}\hat{d}_{\ell m} H_\ell^T +\frac{8\pi M^{-\ell-1}}{2\ell+1}\hat{I}_{\ell m} H_\ell^R \nonumber\\
    &+\left(\omega M\right)^2  P_\ell^{+} +{\cal O}\left(\left(\omega M\right)^3\right).
\end{align}
Here, the coefficients of $P_\ell^{+T}$ and $P_\ell^{+R}$ in $P_\ell^{+}$ are unspecified yet. To determine them, we exploit the resummation originally proposed by Ref.~\cite{HegadeKR:2024agt}: notice from Eqs.~\eqref{eq:TidalMoment} and~\eqref{eq:matching} that
\begin{align}
    \frac{8\pi M^\ell }{2\ell+1}\left(\omega M\right)^2\hat{d}_{\ell m} =&\left(\omega M\right)^2\mathbb{E}_{\ell m}^{(0)}+{\cal O}\left(\left(\omega M\right)^3\right),\nonumber\\
 \frac{8\pi M^{-\ell-1}}{2\ell+1}\left(\omega M\right)^2\hat{I}_{\ell m} =&\left(\omega M\right)^2\mathbb{I}_{\ell m}^{+(0)}+{\cal O}\left(\left(\omega M\right)^3\right),\label{eq:resum}
\end{align}
which allows to express the coefficients of the particular solutions in terms of $\hat{d}_{\ell m}$ and $\hat{I}_{\ell m}$. With the linear relation~\eqref{eq:QFd}, we then find
\begin{eqnarray}
&H&=\frac{8\pi M^\ell }{2\ell+1}\hat{d}_{\ell m} \bigg\{ H_\ell^T+
\left(\omega M\right)^2P_\ell^{+T}\nonumber\\
&+&\frac{2\hat{\cal F}_{\ell}}{{\cal C}^{2\ell+1}}\left[  H_\ell^R +\left(\omega M\right)^2P_\ell^{+R}\right]\bigg\}+{\cal O}\left(\left(\omega M\right)^3\right),\label{eq:H0}
\end{eqnarray}
where we have introduced a stellar compactness,
\beq
{\cal C}:=M/r_0\,. 
\eeq
Here, the tidal response function~$\hat{\cal F}_{\ell}(\omega)$ can be written as
\begin{align}
\hat{\cal F}_{\ell}=&{\cal C}^{2\ell+1}\mathscr{F}_{\ell}^+\left(\omega\right),\label{eq:TidalF}
\end{align}
where we have introduced a {\it rescaled electric-type tidal response function}, 
\begin{widetext}
\begin{align}
    \mathscr{F}_{\ell}^+\left(\omega\right):=& \frac{1}{2} \frac{ \mathbb{I}_{\ell m}^{+(0)}+\omega M \mathbb{I}_{\ell m}^{+(1)}+\left(\omega M\right)^2\mathbb{I}_{\ell m}^{+(2)}}{ \mathbb{E}_{\ell m}^{(0)}+\omega M \mathbb{E}_{\ell m}^{(1)}+\left(\omega M\right)^2\mathbb{E}_{\ell m}^{(2)}} =\frac{1}{2}\left[\frac{\mathbb{I}_{\ell m}^{+(0)}}{\mathbb{E}_{\ell m}^{(0)}}+\left(\omega M\right)\frac{\mathbb{E}_{\ell m}^{(0)}\mathbb{I}_{\ell m}^{+(1)}-\mathbb{E}_{\ell m}^{(1)}\mathbb{I}_{\ell m}^{+(0)}}{\left(\mathbb{E}_{\ell m}^{(0)}\right)^2}\right.\nonumber  \\
    &\left. \quad \quad + \left(\omega M\right)^2\frac{\left(\mathbb{E}_{\ell m}^{(0)}\right)^2 \mathbb{I}_{\ell m}^{+(2)}-\mathbb{E}_{\ell m}^{(0)}\mathbb{E}_{\ell m}^{(1)}\mathbb{I}_{\ell m}^{+(1)}-\mathbb{E}_{\ell m}^{(0)}\mathbb{E}_{\ell m}^{(2)}\mathbb{I}_{\ell m}^{+(0)}+\left(\mathbb{E}_{\ell m}^{(1)}\right)^2\mathbb{I}_{\ell m}^{+(0)}}{\left(\mathbb{E}_{\ell m}^{(0)}\right)^3}\right] + \mathcal{O}\left(\left(\omega M\right)^3\right).
\label{eq:response}
\end{align}
\end{widetext}

\section{Calibration of a tidal response function}\label{Sec:Calibration}
In this section, we first argue that a tidal response function is not uniquely specified; nonetheless, the ambiguity has no impact on observables, provided these are consistently extracted from gravitational waveforms. What we propose, to meaningfully compare response functions, is calibration of a tidal response function, allowing us to determine tidal responses for any compact objects in a unified and simple manner. The parallel discussion in the odd-parity sector is provided in Appendix~\ref{Appendix:CalibrationOdd}.

\subsection{Impact of different choices of $I_\ell^{+ij}$}\label{Sec:Iimpact}
There are four antiderivatives~$I_\ell^{+ij}$ in Eq.~\eqref{eq:H0} as the functional form of the particular solutions~\eqref{eq:Peven} is not uniquely specified, directly leading to the ambiguity of the tidal response function. Then, one might wonder if this ambiguity affects observables, such as gravitational waves being emitted from a tidally interacting spacetime.

The answer is no: observables are unambiguous. The tidally deformed metric is independent of the determination of particular solutions, once an inner boundary condition for a tidally perturbed metric is imposed. This can be understood as follows: the reduced linearized Einstein equation~\eqref{eq:LinearEveneq} is a second-order differential equation, and hence, the general solution has two integration constants. Imposing an inner boundary condition determines one of the two. The remaining integration constant corresponds to the overall factor,~$\mathbb{E}_{\ell m}^{(0)}$, implying that the integration constants of $I_\ell^{+ij}$ are redundant. The degrees of freedom to choose $I_\ell^{+ij}$ correspond to how to decompose the integration constant determined at an inner boundary into external tidal and induced response pieces. 

Observables are thus free from the ambiguity of a tidal response function. Therefore, any choices of the definition of a tidal response function are allowed. However, in practice, it should be noted the following: the difference arising from the implicit assumption on the determination of particular solutions in the definition of tidal response functions gives rise to different values, and then, one would misinterpret them as physically meaningful observations if one does not properly account for the difference between the arbitrarily defined tidal response function and the tidal parameters incorporated in waveform models. This could potentially lead to bias in constraining the nuclear matter equation of state and gravitational theories with gravitational-wave observations. 

\subsection{Calibration}\label{subsec:Calibration}
Calibration of a tidal response function is performed through the following two steps:~(i) fixing the functional form of $P_\ell^+$ in Eq.~\eqref{eq:Peven}, which determines the value of induced multipole moments uniquely;~(ii) exploiting the remaining degrees of freedom to redefine tidal moments while preserving the tidally induced multipole moments to be invariant.

\subsubsection{Determination of the particular solutions}
Although the nature of the tidally deformed object is not yet assumed to be either a BH or a relativistic star, we consider the asymptotic expansion of $fP_\ell^+$, where the factor~$f$ comes from the factorization of the metric perturbation~\eqref{hmunueven}, around $r=2M$,\footnote{The surface~$r=2M$ has physical significance only in the Schwarzschild BH case; in non-BH cases, this expansion corresponds to the analytic continuation of the exterior solution into the interior of a stellar surface. In this sense, the calibration proposed in this section, especially the determination of the particular solutions, privileges BHs. This is also the case in the ``standard" definition for static tidal response based on the analytic properties of hypergeometric functions~(see Refs.~\cite{Gralla:2017djj,Poisson:2020vap}). }
\begin{align}
  & f P_\ell^+ \sim \mathbb{E}_{\ell m}^{(0)}A_\ell^+ \ln f+\mathbb{I}_{\ell m}^{+(0)}  \left[B_\ell^+ \left(\ln f\right)^2+D_\ell^+\ln f\right] \nonumber\\
   &+C_\ell^{+}+2^{2(\ell+1)}\left(\mathbb{E}_{\ell m}^{(0)}\mu_\ell^{T}+\mathbb{I}_{\ell m}^{+(0)}\mu_\ell^{+R}\right)+{\cal O}\left(f\right),
\end{align}
where $A_\ell^+$, $B_\ell^+$, and $D_\ell^+$ are ${\cal O}(1)$ specified constants~(not relevant to the later discussion); $C_\ell^+$ is an unspecified constant part at subleading order to logarithmic terms in the asymptotic expansion of $\mathbb{E}_{\ell m}^{(0)}I_\ell^{+TT} +\mathbb{I}_{\ell m}^{+(0)}I_\ell^{+TR}$ around $r=2M$; $\mu_\ell^{T}$ and $\mu_\ell^{+R}$ are given by 
\begin{align}
     \mu_\ell^{T}=&\frac{\left(\ell+2\right)!\ell!\left(\ell-1\right)!\left(\ell-2\right)!}{2\left(2\ell+1\right)!\left(2\ell-1\right)!},\label{eq:muellT}
\end{align}
while $\mu_\ell^{+R}$ for the lowest $\ell$ values are given by
\begin{align}
\mu_2^{+R}=&-\frac{1}{96} ,\nonumber\\ 
     \mu_3^{+R}=&-\frac{1}{96}, \\ 
     \mu_4^{+R}=&-\frac{59}{15360}.\nonumber
\end{align}
We now require that {\it the asymptotic expansion of $f P_\ell^{+}$ around $r=2M$ does not contain any constant terms at subleading order to the logarithmic terms.} This condition yields
\begin{align}
\label{eq:Cplus}
  C_\ell^{+}=-2^{2\left(\ell+1\right)}\left(\mathbb{E}_{\ell m}^{(0)}\mu_\ell^{T}+\mathbb{I}_{\ell m}^{(0)}\mu_\ell^{+R}\right).
\end{align}
This determines the integration constants of $I_\ell^{+TT}$ and $I_\ell^{+TR}$, specifying the coefficient of the $\bar{r}^{-\ell-1}$ term in the large-distance asymptotic expansions of $P_\ell^{+T}$ and $P_\ell^{+R}$ in Eq.~\eqref{eq:Peven}. We further demand that {\it the asymptotic expansion of $P_\ell^+$ at large distances contains no terms proportional to $\bar{r}^{\ell}$}, determining the integration constants of $I_\ell^{+RT}$ and $I_\ell^{+RR}$, and thereby, specifying the functional form of the particular solutions. When $\kappa_\ell^{(0)}=0$ as in the BH case, the first condition is sufficient for the unique determination of dTLNs; otherwise, the second is also required.

Let us now mention relevant work in this context~\cite{Poisson:2020vap,HegadeKR:2024agt}; it appears that Eq.~(5.55a) and the particular solutions in Appendix~F of Ref.~\cite{Poisson:2020vap} implicitly assume a certain functional form of the particular solution, which leads to an identical tidal response function to ours in the Schwarzschild BH case as will be seen in Sec.~\ref{Sec:dynamicaltide}. One can also choose different forms such as the normalization performed in Ref.~\cite{HegadeKR:2024agt}. As stated before, observables are free from the difference of approaches, provided that the appropriate definition of the tidal response function is used to evaluate the dynamics. There is no ``correct'' definition; we now suggest a simple definition for the response, that allows one to recover the well-known results of vanishing TLNs for BHs~\cite{Binnington:2009bb,Damour:2009vw,Poisson:2020vap}.

\subsubsection{Redefinition of tidal moments}
Let us redefine tidal moments at the non-adiabatic regime as follows~\cite{Poisson:2020vap,Pitre:2023xsr}:
\begin{align}
    \bar{\mathbb{E}}_{\ell m}^{(1)}=\mathbb{E}_{\ell m}^{(1)}-\xi_{\ell m}^{+(1)},~~\bar{\mathbb{E}}_{\ell m}^{(2)}= \mathbb{E}_{\ell m}^{(2)}-\xi_{\ell m}^{+(2)},\label{eq:RedefinitionOfTidalMoments}
\end{align}
where $\xi_{\ell m}^{+(1)}$ and $\xi_{\ell m}^{+(2)}$ are not necessarily infinitesimal. Then, induced multipole moments remain invariant if and only if TDNs and dTLNs are also redefined:
\begin{align}
    \bar{\nu}_\ell^{(0)}=&\nu_\ell^{(0)}-i\kappa_\ell^{(0)}\frac{\xi_{\ell m}^{+(1)}}{\mathbb{E}_{\ell m}^{(0)}},\\
    \bar{\kappa}_\ell^{(1)}=&\kappa_\ell^{(1)}+i \nu_{\ell}^{(0)}\frac{\xi_{\ell m}^{+(1)}}{\mathbb{E}_{\ell m}^{(0)}}-\kappa_\ell^{(0)}\left[\frac{\mathbb{E}_{\ell m}^{(1)}-\xi_{\ell m}^{+(1)}}{\left(\mathbb{E}_{\ell m}^{(0)}\right)^2}\xi_{\ell m}^{+(1)}-\frac{\xi_{\ell m}^{+(2)}}{\mathbb{E}_{\ell m}^{(0)}}\right].
\end{align}

Choosing $\xi_{\ell m}^{+(1)}=\mathbb{E}_{\ell m}^{(1)}$ and $\xi_{\ell m}^{+(2)}=\mathbb{E}_{\ell m}^{(2)}$, one can {\it remove the contributions of tidal moments at the non-adiabatic regime}. We thus obtain the calibrated tidal response function in vacuum GR,
\begin{align}
\hat{\cal \tilde{F}}_\ell\left(\omega\right):= \kappa_\ell^{(0)}+i\omega M \tilde{\nu}_\ell^{(0)}+\left(\omega M\right)^2 \tilde{\kappa}_\ell^{(1)}+{\cal O}\left(\omega^3\right),\label{eq:CTRF}
\end{align}
with
\begin{align}
    \tilde{\nu}_\ell^{(0)}:=&\nu_\ell^{(0)}-i\kappa_\ell^{(0)}\frac{\mathbb{E}_{\ell m}^{(1)}}{\mathbb{E}_{\ell m}^{(0)}},\label{eq:redefinitionTDNs}\\
    \tilde{\kappa}_\ell^{(1)}:=&\kappa_\ell^{(1)}+i \nu_{\ell}^{(0)}\frac{\mathbb{E}_{\ell m}^{(1)}}{\mathbb{E}_{\ell m}^{(0)}}+\kappa_\ell^{(0)}\frac{\mathbb{E}_{\ell m}^{(2)}}{\mathbb{E}_{\ell m}^{(0)}}.\label{eq:redefinitiondTLNs}
\end{align}
The above redefinition simplifies the expression for the tidally induced multipole moments via the linear relation~\eqref{eq:QFd}:
\begin{align}
\hat{I}_{\ell m}\left(\omega\right)=&2r_0^{2\ell+1}\hat{\cal \tilde{F}}_\ell \left(\omega\right)\hat{d}_{\ell m}^{(0)},
\end{align}
with $\hat{d}_{\ell m}^{(0)}\equiv \hat{d}_{\ell m}(\omega=0)$. Notice that $\tilde{\nu}_\ell^{(0)}=\nu_\ell^{(0)}$ when $\kappa_\ell^{(0)}=0$, which is the case for the Schwarzschild BH as we demonstrate in the next section.

\subsection{Advantages of the calibrated tidal response function}
The calibration~\eqref{eq:CTRF}~(and~\eqref{eq:CTRFodd} for the odd-parity sector) leads to vanishing dTLNs of the Schwarzschild BH in both sectors, as will be demonstrated in Sec.~\ref{Sec:dynamicaltide}. This property allows one to easily define the values for relativistic stars, such as neutron stars, as the differences from the BH values in a unified manner. By reinterpreting the calibrated tidal response function as the difference between the non-BH value and the BH value of the same mass, the tidal deformability parameter incorporated in the gravitational-waveform modeling,~$\Lambda^{\rm GW}$, is related to that computed from the calibrated value,~$\tilde{\Lambda}$, as $\Lambda^{\rm GW}-\Lambda^{\rm GW}_{\rm (BH)}=\tilde{\Lambda}$, where $\Lambda^{\rm GW}_{\rm (BH)}$ is the tidal deformability parameter for the BH in the PN waveform. Once $\Lambda^{\rm GW}_{\rm (BH)}$ is accurately computed from binary BH merger simulations using the PN waveform model, one can determine $\tilde{\Lambda}$ from $\Lambda^{\rm GW}$ measured with observations, allowing constraints on the nuclear matter equation of state. We stress that, in any case, the definition used in computation should be noted together with the resulting tidal response functions to avoid confusion.

\subsection{Alternative definition}\label{sec:alternative}
Depending on the context, one can adopt other definitions of tidal response with other advantages~\cite{HegadeKR:2024agt,Katagiri:2024fpn}. For instance, the normalization proposed in Ref.~\cite{HegadeKR:2024agt} is a different procedure to fix particular solutions, demanding that the large-distance asymptotic expansion of particular solutions does not contain any terms proportional to $\bar{r}^{\ell}$ and $\bar{r}^{-\ell-1}$, where $\bar{r}$ is a harmonic radial coordinate. The only difference from our particular solutions lies in the coefficient of the $\bar{r}^{-\ell-1}$ term. As will be seen in Sec.~\ref{sec:HRY}, the dTLNs of the Schwarzschild BH are no longer zero, although we stress again that this does not indicate a disagreement in the prediction of observables.

The ambiguity in the choice of particular solutions emerges even in static TLNs beyond vacuum GR in a perturbative framework. This alternative approach to determine the particular solutions is more useful in non-vacuum and/or non-GR setups~\cite{Katagiri:2024fpn}.

\section{Dynamical tides of a Schwarzschild black hole}\label{Sec:dynamicaltide}
As an application of the calibrated tidal response function~\eqref{eq:CTRF}, we now consider Schwarzschild BHs. We first demonstrate that a ``bare" tidal response function~\eqref{eq:TidalF} of the Schwarzschild BH depends on  $C_\ell^{+}$. The technical details of computation of the tidal response functions in both even-parity and odd-parity sectors are provided in Appendix~\ref{Appendix:TidalResponseFunctions}. We then show that the Schwarzschild BH dTLNs vanish at any multipole order within the calibration. Additionally, we briefly discuss the extension of the calibrated tidal response function to ${\cal O}(\omega^3M^3)$, and then, analyze the next-to-leading dissipative tidal effect.

\subsection{Bare tidal response function}\label{Sec:TidalResponseFunction}
A bare tidal response function is determined (while leaving the particular solutions ambiguous) by imposing an inner boundary condition on the metric perturbation. In the BH case, the corresponding condition is an ingoing-wave condition at the BH horizon. We thus find 
\begin{align}
    \kappa_\ell^{(0)}=&0, \label{eq:zeroFp0}\\
    \nu_\ell^{(0)}=&\frac{\left(\ell+2\right)!\ell!\left(\ell-1\right)!\left(\ell-2\right)!}{2\left(2\ell+1\right)!\left(2\ell-1\right)!},\label{eq:TDNs}\\
    \kappa_\ell^{(1)}=&\frac{4\ell\left(\ell+1\right)\left(\gamma+\psi\left(\ell\right)\right)-\ell\left(5\ell+1\right)-4}{\ell \left(\ell+1\right)}\nu_\ell^{(0)}\nonumber\\
    &-\frac{C_\ell^{+}}{2^{2\ell+2}\mathbb{E}_{\ell m}^{(0)}}.\label{eq:dTLNs}
\end{align}
Here, $\nu_\ell^{(0)}$ is identical to $\mu_\ell^{T}$ in Eq.~\eqref{eq:muellT}.

Equation~\eqref{eq:zeroFp0} recovers the well-known vanishing TLNs of the Schwarzschild BH~\cite{Binnington:2009bb,Poisson:2020vap,Poisson:2021yau}. The TDNs, $\nu_\ell^{(0)}$, are in perfect agreement with other independent analytical results~\cite{Poisson:2020vap,Chia:2020yla,Katagiri:2023yzm,Chakraborty:2023zed}. We obtain the same values of $\kappa_\ell^{(0)}$ and $\nu_\ell^{(0)}$ in the odd-parity sector as well in Appendix~\ref{Appendix:TidalResponseFunctions}. As mentioned earlier, $\kappa_\ell^{(0)}$ and $\nu_\ell^{(0)}$ do not change under the redefinition of tidal moments because $\kappa_\ell^{(0)}=0$. On the other hand, the expression~\eqref{eq:dTLNs} demonstrates that the value of the Schwarzschild BH dTLNs depends on the choice of $C_\ell^{+}$. Note that the metric perturbation with the assignment of TLNs~\eqref{eq:zeroFp0}, TDNs~\eqref{eq:TDNs}, and dTLNs~\eqref{eq:dTLNs} is smooth at the BH horizon in ingoing Eddington-Finkelstein coordinates, which is supported by the analysis of Ref.~\cite{Poisson:2020vap}.

The ingoing-wave boundary condition determines the tidal moments at higher order as well, thereby obtaining 
\begin{align}
\label{eq:E1}
    \mathbb{E}_{\ell m}^{(1)}=4i\frac{\ell\left(\ell+1\right)\left(\psi\left(\ell\right)+\gamma \right)-\ell^2-1}{\ell\left(\ell+1\right)} \mathbb{E}_{\ell m}^{(0)}.
\end{align}
Here, $\psi(\ell)$ and $\gamma (\simeq 0.57721)$ are a digamma function and Euler's constant, respectively. The coefficient~$\mathbb{E}_{\ell m}^{(2)}$ can also be determined though we do not need it for computing dTLNs when TLNs vanish.

\subsection{Zero dTLNs within the calibration}\label{subsec:Calibration}
 First, the condition~\eqref{eq:Cplus} takes the form~(note $\mathbb{I}_{\ell m}^{+(0)}=0$),
\begin{align}
    C_\ell^{+}=-2^{2\left(\ell+1\right)}\mathbb{E}_{\ell m}^{(0)}\nu_\ell^{(0)},
\end{align}
where we have used $\mu_\ell^{T}=\nu_\ell^{(0)}$. Then, Eq.~\eqref{eq:dTLNs} with Eq.~\eqref{eq:E1} leads to
\begin{align}
\kappa_{\ell}^{(1)}=-i \frac{\mathbb{E}_{\ell m}^{(1)}}{\mathbb{E}_{\ell m}^{(0)}}\nu_\ell^{(0)}\label{eq:calibrateddTLNs}.
\end{align}

Next, we use the redefinition in Eqs.~\eqref{eq:redefinitionTDNs} and~\eqref{eq:redefinitiondTLNs}. Since $\kappa_\ell^{(0)}$ vanishes as shown in Eq.~\eqref{eq:zeroFp0}, the value of $\nu_\ell^{(0)}$ is invariant under the transformation but dTLNs shift, thereby obtaining
\begin{align}
    \tilde{\kappa}_\ell^{(1)}=0.
\end{align}
This shows the vanishing of dTLNs for Schwarzschild BHs with arbitrary~$\ell$, proving the conjecture by Poisson~\cite{Poisson:2020vap}. In the same manner, we show that dTLNs are zero also in the odd-parity sector in Appendix~\ref{Appendix:TidalResponseFunctions}.

\subsection{Nonzero dTLNs within the alternative definition}\label{sec:HRY}
If one applies the alternative definition by Hegade, Ripley, and Yunes~\cite{HegadeKR:2024agt}, stated in Sec.~\ref{sec:alternative}, the dTLNs of the Schwarzschild BH, denoted as HRY, are no longer zero:
\begin{align}
    \kappa_2^{{\rm HRY},(1)}=-\frac{32}{1575},~~ \kappa_3^{{\rm HRY},(1)}=\frac{109}{352800}.
\end{align}
Again, this does not indicate discrepancies in the predictions of observables. We find different values simply due to different conventions.

\subsection{Extension to the next-to-leading dissipative tidal effect}
Let us briefly discuss the extension of the calibration~\eqref{eq:CTRF} to ${\cal O}(\omega^3M^3)$. The expansion of the radial component of the metric perturbation, say $H^{(3)}(r)$, satisfies an inhomogeneous equation of the form of Eq.~\eqref{eq:eqforH2} but sourced by $H^{(1)}$:
\begin{align}
    {\cal L}_\ell^+\left[H^{(3)}\right]={\cal S}_\ell^+\left[H^{(1)}\right].
\end{align}
Notice that the general solution of $H^{(3)}$ takes the same form as $H^{(2)}$. 

The bare tidal response function~\eqref{eq:TidalF} captures the next-to-leading dissipative tidal effect that is quantified by $\nu_\ell^{(1)}$ with
\begin{align}
    \hat{\cal F}_{\ell}(\omega)=&\kappa_{\ell}^{(0)}+i\omega M\nu_{\ell}^{(0)}+\left(\omega M\right)^2 \kappa_{\ell}^{(1)}\nonumber\\
    &-i (\omega M)^3\nu_\ell^{(1)}+{\cal O}(\omega^4)\,.
\end{align}
The value of $\nu_\ell^{(1)}$ depends on the functional form of the particular solutions in the general solution of $H^{(3)}$, which is fixed in the parallel manner to Eq.~\eqref{eq:Cplus}. We then find the higher-order corrections to the redefinition of the tidal moments and the calibrated tidal response function in Sec.~\ref{Sec:Calibration}:
\begin{align}
    \bar{\mathbb{E}}_{\ell m}^{(3)}=& \mathbb{E}_{\ell m}^{(3)}-\xi_{\ell m}^{+(3)},
\end{align}
and
\begin{align}
\hat{\cal \tilde{F}}_\ell \left(\omega\right)=& \kappa_\ell^{(0)}+i\omega M \tilde{\nu}_\ell^{(0)}+\left(\omega M\right)^2 \tilde{\kappa}_\ell^{(1)}\nonumber\\
&-i (\omega M)^3 \tilde{\nu}_\ell^{(1)}+{\cal O}\left(\omega^4\right),\label{Eq:CTRFhigher}
\end{align}
with
\begin{align}
    \tilde{\nu}_\ell^{(1)}:=&\nu_\ell^{(1)}+i \kappa_\ell^{(1)} \frac{\xi_{\ell m}^{+(1)}}{\mathbb{E}_{\ell m}^{(0)}} \nonumber\\
    &-\nu_\ell^{(0)}\frac{\left(\xi_{\ell m}^{+(1)}\right)^2-\mathbb{E}_{\ell m}^{(1)} \xi_{\ell m}^{+(1)}+\mathbb{E}_{\ell m}^{(0)} \xi_{\ell m}^{+(2)}}{\left(\mathbb{E}_{\ell m}^{(0)}\right)^2}\nonumber\\
    &+i \frac{\kappa_\ell^{(0)}}{\left(\mathbb{E}_{\ell m}^{(0)}\right)^3} \left[\left(\xi_{\ell m}^{+(1)}\right)^3+\left(\mathbb{E}_{\ell m}^{(1)}\right)^2 \xi_{\ell m}^{+(1)}\right. \\
    &-\mathbb{E}_{\ell m}^{(0)}\xi_{\ell m}^{+(1)}\left(\mathbb{E}_{\ell m}^{(2)}-2\xi_{\ell m}^{+(2)}\right)\nonumber\\
    &\left.-\mathbb{E}_{\ell m}^{(1)}\left\{\mathbb{E}_{\ell m}^{(0)}\xi_{\ell m}^{+(2)}+2\left(\xi_{\ell m}^{+(1)}\right)^2 \right\}+\left(\mathbb{E}_{\ell m}^{(0)}\right)^2\xi_{\ell m}^{+(3)}\right].\nonumber
\end{align}
The requirement of no outgoing waves from the BH horizon determines the BH values of the calibrated tidal response function~\eqref{Eq:CTRFhigher}, e.g.,
\begin{align}
    \tilde{\nu}_2^{(1)}=&\frac{17+3\pi^2}{135},~~\tilde{\nu}_3^{(1)}=\frac{2+3\pi^2}{3780}.
\end{align}
To the best of our knowledge, this is the first computation of the next-to-leading dissipative tidal effect for Schwarzschild BHs.

\section{Logarithmic running corrections}\label{Sec:Running}
We discuss the physical interpretation of logarithmic corrections appearing in the asymptotic expansion of the body zone metric at large distances. One possible interpretation for the origin of some of them is the deviation of light cones~$t \pm \bar{r}$ associated with the~$(t,\bar{r},\theta,\varphi)$~coordinate system from the true light cone through which a tidal perturbation propagates~\cite{Blanchet:1985sp,Blanchet:1986dk}. This deviation is expressed by the logarithmic term, $2M \ln(\bar{r}/2M-1/2)$, as known in the context of the multipolar post-Minkowskian-PN formalism~\cite{Blanchet:1985sp,Blanchet:1986dk,Blanchet:2013haa,Blanchet:2024mnz}. As a result of the deviation, tidal perturbations climbing the gravitational well in the body zone experience phase shifts with respect to propagation in the PN zone. 

The emergence of logarithmic terms in perturbations also often involve wave-propagation effects, i.e., tail effects~\cite{Poisson:1993vp,Blanchet:1985sp,Blanchet:1986dk}. The literature~\cite{Sasaki:1994rw,Tagoshi:1994sm,Sasaki:2003xr} show the emergence of logarithmic terms at ${\cal O}(\omega^2M^2)$ in the PN expansion of gravitational perturbations on a Schwarzschild background, which involve corrections first at $3$PN order in the waveform. The corrections are thought as tails of tails, which are caused by the backscattering of tails of the gravitational perturbation due to spacetime curvature~\cite{Blanchet:1997jj}. At $1.5$PN order, the waveform is corrected by the tails that are caused by the backscattering of gravitational perturbations due to spacetime curvature at large distances from the body~\cite{Poisson:1993vp,Tagoshi:1994sm,Sasaki:2003xr,Blanchet:1997jj}, which should be captured in Eq.~\eqref{eq:H0} up to ${\cal O}(\omega M)$, but the extraction of the information from Eq.~\eqref{eq:H0} is beyond the scope of our analysis.

We found a logarithmic term at the same order as tidally induced multipole moments, consistent with findings in Refs.~\cite{Chakrabarti:2013lua,Saketh:2023bul,Perry:2023wmm,Chia:2024bwc}, 
\begin{align}
k_\ell^u=-\nu_\ell^{(0)}\left(2M \omega\right)^2\ln \left(\frac{2M}{\bar{r}}\right).\label{eq:logrunning}
\end{align}
Here, $\nu_\ell^{(0)}$ is given by Eq.~\eqref{eq:TDNs}. One might expect that the phase shift and the multiple backscattering of tidal perturbations in the buffer zone induce this logarithmic deformation of the tidally induced multipole moments.\footnote{ 
One might think that this picture is unlikely to hold in spacetimes which do not cause tails, such as asymptotically (anti-) de Sitter spacetimes~\cite{Ching:1994bd,Ching:1995tj}. We emphasize that a local geometry around the central object in a small scale compared to the curvature radius of the external spacetime is relevant in describing tidal interactions in binary systems. Within a low-frequency regime, the geometry around the body is well approximated by an asymptotically flat spacetime~\cite{Cardoso:2004hs,Uchikata:2011zz,Katagiri:2020mvm}, implying that the body admits tails within a certain stage of its time evolution. Indeed, for example, the numerical analysis in Ref.~\cite{Cardoso:2015fga} shows that a small Schwarzschild BH in anti-de Sitter spacetime displays power-law tails in the early stage of the ringdown signal.} In the current analysis, however, it is difficult to identify the origin of the individual logarithmic terms in the asymptotic expansion of the tidally perturbed metric.

The worldline EFT framework~\cite{Kol:2011vg,Saketh:2023bul,Ivanov:2024sds} indicates that the logarithmic correction results in cutoff-radius dependence of dTLNs in the matching scheme with a point-particle description. This dependence arises from the ambiguous decomposition of the gravitational field of the tidally deformed body and its tidal environment, requiring a cutoff radius to isolate them. The dTLNs are thus scale dependent and are altered by varying the cutoff radius. It would be beneficial to conduct a detailed study of this logarithmic correction within the PN framework, including a comparison with the EFT framework.

\section{Summary and discussion}\label{Sec:discussion}
In this paper, we highlighted subtleties in modeling relativistic dynamical tides and proposed calibration of a tidal response function, allowing one to determine tidal responses for any compact objects in a unified and simple manner. Our analysis relies on a low-frequency expansion of the linearized Einstein equations in vacuum and solves the perturbation equations order by order. The tidally perturbed metric is then asymptotically matched with a PN metric, assigning tidal moments and tidally induced multipole moments to the integration constants in the perturbed metric. Our argument and analysis can be generalized to spinning BHs and stars. 

The ambiguity in the decomposition of the metric perturbation into external tidal and response pieces leads to an arbitrary definition of tidal response functions. Nonetheless, for a given external source, a tidally deformed geometry is completely specified, irrespective of the definition of the tidal response function, once one imposes inner boundary conditions; therefore, the description of any dynamics on the deformed background is unambiguous. In practice however, implicit assumptions in the definition of tidal responses could lead to a bias in constraining dTLNs or the nuclear matter equation of state if one does not properly account for the differences between the arbitrarily defined tidal response functions and the tidal parameters incorporated in waveform models. Within the worldline EFT approach in Refs.~\cite{Saketh:2022xjb,Saketh:2024juq}, the dissipative tidal coefficients are linked to observables, such as the evolution of mass and spin of objects as well as the waveform, thereby allowing one to unambiguously evaluate the dissipative tidal effect. It is beneficial to relate TDNs calibrated within our framework to the change of mass of a tidally deformed body in the similar manner. Note that, although the BH TDNs in Eq.~\eqref{eq:TDNs} remain invariant under the redefinition introduced in Sec.~\ref{Sec:Calibration}, this is not generically the case for relativistic viscous stars. However, the redefinition does not alter the underlying physics, as the TDNs in the current framework are not necessarily linked to invariant quantities, such as changes in mass. Nevertheless, it is worth mentioning that the BH TDNs in Eq.~\eqref{eq:TDNs} are consistent with those derived in the EFT approach~\cite{Saketh:2022xjb,Saketh:2024juq}, apart from the conventional numerical factor. 

The calibration, especially the determination of particular solutions, privileges GR BHs in a definition for tidal response functions that are supposed to be applicable to any compact objects. Consequently, the calibrated tidal response function recovers the known results for the Schwarzschild BH~(vanishing TLNs~\cite{Damour:2009vw,Binnington:2009bb,Poisson:2020vap,Poisson:2021yau} and the analytic expression for TDNs~\cite{Poisson:2020vap,Chia:2020yla,Katagiri:2023yzm,Chakraborty:2023zed}) and leads to vanishing of the dTLNs of the Schwarzschild BH at any multipole order in even and odd parities. To determine the dynamical tides of neutron stars from observations, one could use the same calibrated tidal response function, and then, reinterpret it as the difference between the BH value and the values of a neutron star with the same mass, following a previous suggestion~\cite{Gralla:2017djj}. One can obtain the value of the calibrated tidal response function by subtracting a PN-defined BH dTLN value (extracted from highly accurate binary BH simulations using a PN waveform model) from a measured value for a neutron star from a binary neutron star merger observation using the PN waveform. This allows one to translate it into information on the nuclear matter equation of state used in computations in the body zone. 

As the worldline EFT framework~\cite{Kol:2011vg,Saketh:2023bul,Ivanov:2024sds} indicates, the value of dTLNs is scheme dependent due to the presence of the logarithmic running correction discussed in Sec.~\ref{Sec:Running}. Specifically, it is altered by varying the cutoff scale in the matching scheme with a point-particle description. We stress that the vanishing of the dTLNs of a Schwarzschild BH, shown in this work, is also scheme dependent, which allows us to compare the dTLNs of non-BH objects in a unified manner. Our calibration corresponds to choosing a specific scale in the EFT approach. What matters is that the definition of dTLNs varies, and that does not immediately indicate inconsistencies so long as they are properly linked to observables.

Let us comment on some limitations in our framework.
The framework for matching with a PN metric is performed at the leading Newtonian order only and assumes that the tidal potential of a BH can be approximated by a Newtonian potential. This can be justified when the orbital separation is sufficiently large compared to the radii of the binary constituents. Indeed, our framework was able to recover the vanishing TLNs for a Schwarzschild BH which was also derived in a fully relativistic setup in Ref.~\cite{Poisson:2021yau}. It is also expected that the above approximation is still valid in a certain dynamical regime. Additionally, the harmonic coordinate system used in this work is not exactly harmonic in the tidally perturbed spacetime but is still a good approximation in a certain inspiral stage. It would be beneficial for better understanding to analyze the dynamical tidal properties of a Schwarzschild BH in a fully relativistic setting as in Ref.~\cite{Poisson:2021yau} and compare the full PN order results with our Newtonian order results.

We found logarithmic running corrections to dTLNs identical to those in Refs.~\cite{Chakrabarti:2013lua,Saketh:2023bul,Perry:2023wmm,Chia:2024bwc}. One might expect some connection with tail effects. However, this remains speculative and is not supported by any robust analysis, except for the comparison with the classical results from BH perturbation theory and multipolar post Minkowskian and post Newtonian formalism~\cite{Poisson:1993vp,Tagoshi:1994sm,Sasaki:2003xr,Blanchet:1997jj}. The worldline EFT framework~\cite{Kol:2011vg,Saketh:2023bul,Ivanov:2024sds} indicates that the logarithmic term results in a cutoff-radius dependence of dTLNs, where the gravitational field of a tidally deformed body is isolated from its tidal environment. How this EFT perspective is incorporated into the current PN matching framework remains unclear. A comprehensive analysis of this logarithmic correction within the PN framework, including a comparison with the EFT approach, would be beneficial.

We note the ambiguity in the so-called near-zone approximation often used in literature. For example, one can still add terms that were dropped from the radial Teukolsky equation into Eq.~(4.2) of Ref.~\cite{Perry:2023wmm} under low-frequency and near-horizon assumptions, implying that the analysis therein neglects some relevant terms in a low-frequency expansion. The resulting tidal response function therefore can differ from results of an accurate and consistent perturbative expansion in a subjective way. In fact, taking the constant,~$k_E$, in Eq.~(4.12) of Ref.~\cite{Perry:2023wmm} into account, the~$r^{-\ell-1}$ term in Eq.~(4.13) therein acquires a non-vanishing constant coefficient but its value would be altered if one adopts other near-zone approximation schemes as in Refs.~\cite{Chakraborty:2023zed,Bhatt:2024yyz}. Thus, these approximations require further careful consideration with robust formulation. Our current work corresponds to a complementary analysis to the above works in the Schwarzschild BH case. An extension of our analysis to rotating spacetime backgrounds is left for future work.

The ambiguity of tidal response functions can arise in general in a perturbative framework, for example, in the context of alternative theories of gravity under the small deviation assumption from GR. Hence, the definition of a tidal response function should be carefully taken into account together with computed values in such a framework. We propose a canonical definition of tidal response functions in a perturbative framework and provide BH tidal deformabilities in various alternative theories of gravity in our companion work~\cite{Katagiri:2024fpn}.

\acknowledgments
We are indebted to Eric Poisson and Hiroyuki Nakano for helpful comments and insights. We are grateful to Abhishek Hegede and Nicolas Yunes for carefully reading the manuscript and providing us useful feedback. We thank Zihan Zhou for providing useful information on relevant papers based on his insights into the renormalization-group flow of dTLNs with the EFT approach and Muddu Saketh for informing us about relevant important papers.  We acknowledge support by VILLUM Foundation (grant no.\ VIL37766) and the DNRF Chair program (grant no.\ DNRF162) by the Danish National Research Foundation.
V.C.\ acknowledges financial support provided under the European Union’s H2020 ERC Advanced Grant “Black holes: gravitational engines of discovery” grant agreement no.\ Gravitas–101052587. 
Views and opinions expressed are however those of the author only and do not necessarily reflect those of the European Union or the European Research Council. Neither the European Union nor the granting authority can be held responsible for them.
This project has received funding from the European Union's Horizon 2020 research and innovation programme under the Marie Sk{\l}odowska-Curie grant agreement No 101007855 and No 101131233.
K.Y. acknowledges support from NSF Grant No.~PHY-2207349, No.~PHY-2309066, No.~PHYS-2339969, and the Owens Family Foundation.

\appendix

\begin{widetext}

\section{Explicit forms of perturbation equations in the even sector}\label{Appendix:ExplicitForms}
The quantities $W_1^+$ and $W_2^+$ in Eq.~\eqref{eq:LinearEveneq} are:
\begin{align}
    W_1^+:=&W_{10}^++W_{12}^+\omega^2+W_{14}^+\omega^4,\label{eq:P1p}\\
    W_2^+:=&W_{20}^++W_{22}^+\omega^2+W_{24}^+\omega^4+W_{26}^+\omega^6,\nonumber
\end{align}
where
\begin{align}
    W_{10}^+:=&-\frac{2\left(\ell+2\right)\ell}{\sigma^+} f\left(r-M\right)r\left[\left(\ell^2 f-1\right)r+2M  \right],\\
      W_{12}^+:=&\frac{8M r^2}{\sigma^+}\left[2\left\{\left(\ell+1\right)\ell f-1\right\}r^2+7M r-9M^2\right],\\
      W_{14}^+:=&\frac{8r^6}{\sigma^+}\left(r-5M\right),
\end{align}
and
\begin{align}
    W_{20}^+:=&\frac{1}{f \sigma^+} \left[ \ell^2 \left\{\ell^3\left(\ell+3\right)f+\ell^2-3\ell-2  \right\}f^2r^2-2\ell^2\left\{ \ell\left(\ell-3\right)-2
 \right\} Mf^2r+4\left(\ell+2\right)\left(\ell+1\right)\ell\left(\ell-1\right)f^2 M^2 \right],\\
      W_{22}^+:=&-\frac{1}{f \sigma^+} \left[5\ell \left(\ell+2\right)f\left(\ell^2 f-1\right)r^4+2\ell \left(17\ell+22\right)M f r^3 -44\left(\ell+1\right)\ell M^2 f r^2+16M \left(r^3-11M r^2+32M^2 r-27M^3\right)  \right],\\
       W_{24}^+:=&\frac{1}{f \sigma^+} \left[8\left(\ell+1\right)\ell fr^6 -4r^4\left(6r^2-36M r+53M^2  \right)  \right],\\
      W_{26}^+:=&-\frac{4r^8}{f \sigma^+}.
\end{align}
Here, we have defined
\begin{align}
    \sigma^+:=&-2\left(\ell+2\right)\ell M f^2 r^3-\left(\ell+2\right) \ell f^2\left(\ell^2 f-1\right)r^4 -\left[36M^2f r^4-32M f r^5-4\left\{ \left(\ell+1\right)\ell f-2\right\}fr^6\right]\omega^2-4fr^8\omega^4.
\end{align}

The right-hand side of Eq.~\eqref{eq:eqforH2} takes the form,
\begin{align}
\label{eq:Tp}
    {\cal S}_\ell^+:=\frac{1}{\ell\left(\ell^3+2\ell^2-\ell-2\right)f^4}&\left(\alpha^+ r\frac{d}{dr}+\beta^+\right),
\end{align}
with 
\begin{align}
    \alpha^+:=&-8\lambda -8\left(2-7\lambda \right)\left(\frac{M}{r}\right)+128\left(1-\lambda\right)\left(\frac{M}{r}\right)^2-48\left(7-\lambda \right)\left(\frac{M}{r}\right)^3+288\left(\frac{M}{r}\right)^4,\\
    \beta^+:=&-\lambda \ell\left(\ell+1\right)+4\left[\ell\left(\ell+1\right)\left(\ell^2+\ell-4\right)-4\right]\left(\frac{M}{r}\right)-4\left[\left(\ell+4\right)\left(\ell+1\right)\ell\left(\ell-3\right)-36\right]\left(\frac{M}{r}\right)^2\\
    &-48\left(\ell^2+\ell+8\right)\left(\frac{M}{r}\right)^3+288\left(\frac{M}{r}\right)^4.\nonumber
\end{align}
\end{widetext}
Here, we introduced $\lambda=(\ell+2)(\ell-1)$.

\section{Low-frequency expansion}\label{Appendix:LowFrequencyExpansion}
We here solve linearized Einstein equations in vacuum in terms of low-frequency expansions.

\subsection{Even-parity sector}
Equation~\eqref{eq:LinearEveneq} is expanded into Eqs.~\eqref{eq:eqforH0},~\eqref{eq:eqforH1}, and~\eqref{eq:eqforH2} with small~$\omega M$. In what follows, we solve them order by order. 
\subsubsection{Zeroth and first order}
The general solution of Eq.~\eqref{eq:eqforH0} is
 \begin{align}
    \label{eq:generalsolH00} H^{(0)}=&\mathbb{E}_{\ell m}^{(0)}H_\ell^T+\mathbb{I}_{\ell m}^{+(0)} H_\ell^R,
\end{align}
with
\begin{align}
    H_\ell^T=& f\left(\frac{r}{M}\right)^\ell \!~_2F_1\left(-\ell+2,-\ell;-2\ell;2M/r\right),\\
    H_\ell^R=&f\left(\frac{M}{r}\right)^{\ell+1}\!~_2F_1\left(\ell+1,\ell+3;2\ell+2;2M/r\right),
\end{align}
which are identical to Eqs.~\eqref{eq:HT} and~\eqref{eq:HR}, respectively. Here, $\mathbb{E}_{\ell m}^{(0)}$ and $\mathbb{I}_{\ell m}^{+(0)}$ are dimensionless constants.

Equation~\eqref{eq:eqforH1} takes the same form as Eq.~\eqref{eq:eqforH0}. Therefore, the general solution is given by
\begin{align}
        \label{eq:generalsolH01}H^{(1)}=&\mathbb{E}_{\ell m}^{(1)}H_\ell^T+\mathbb{I}_{\ell m}^{+(1)} H_\ell^R,
\end{align}
with dimensionless constants, $\mathbb{E}_{\ell m}^{(1)}$ and $\mathbb{I}_{\ell m}^{+(1)}$. 

\subsubsection{Second order}
Equation~\eqref{eq:eqforH2} is an inhomogeneous equation. With variation of the constant method, the general solution consists of the general solution of a homogeneous equation and a particular solution,
\begin{align}
\label{eq:generalH02}
    H^{(2)}=&\mathbb{ E}_\ell^{(2)}H_\ell^T+\mathbb{I}_{\ell m}^{+(2)} H_\ell^R\nonumber\\
    &-\left(\int\frac{H_\ell^R(r){\cal S}_\ell^+\left[H^{(0)}\right]}{{\cal W}^+}dr\right)H_\ell^T\\
    &+\left(\int\frac{H_\ell^T(r){\cal S}_\ell^+\left[H^{(0)}\right]}{{\cal W}^+}dr\right)H_\ell^R,\nonumber
\end{align}
with
\begin{align}
    {\cal W}^+=-\frac{2\ell+1}{M}f^{-1}\left(\frac{M}{r}\right)^2.
\end{align} 
Here, $\mathbb{E}_{\ell m}^{(2)}$ and $\mathbb{I}_{\ell m}^{+(2)}$ are dimensionless constants. Equation~\eqref{eq:generalH02} can be cast into
\begin{align}
\label{eq:generalsolH02}
    H^{(2)}=&\mathbb{E}_{\ell m}^{(2)}H_\ell^T+\mathbb{I}_{\ell m}^{+(2)} H_\ell^R+P_\ell^+.
\end{align}
Here, we have introduced
\begin{align}
    P_\ell^+=\mathbb{E}_{\ell m}^{(0)} P_\ell^{+T}+\mathbb{I}_{\ell m}^{+(0)} P_\ell^{+R},
\end{align}
with
\begin{align}
P_\ell^{+T}=&I_\ell^{+TT} H_\ell^R- I_\ell^{+RT} H_\ell^T,\\
P_\ell^{+R}=&I_\ell^{+TR} H_\ell^R-I_\ell^{+RR} H_\ell^T,
\end{align}
and 
\begin{align}
  I_\ell^{+ij}\left(r\right):=\int \frac{H_\ell^i {\cal S}_\ell^+\left[H_\ell^j\right]}{{\cal W}^+}dr,
\end{align}
where $i,j$ run $T,R$.

\subsection{Odd-parity sector}

\subsubsection{Linearized Einstein equations in vacuum}
In the Regge-Wheeler gauge~\cite{PhysRev.108.1063}, an odd-parity linear metric perturbation with a spherical harmonic decomposition in the Fourier domain reads
\begin{align}
h_{\mu\nu}^{\rm (odd)}dx^\mu dx^\nu=2h_1 S_{\ell m}^\varphi e^{-i\omega t}dtd\varphi-2h S_{\ell m}^\theta e^{-i\omega t}dtd\theta,
\end{align}
where $(S_{\ell m}^\theta, S_{\ell m}^\varphi):=(-\partial_\varphi Y_{\ell m}/\sin\theta, \sin\theta \partial_\theta Y_{\ell m})$. Here, $h=h(r)$ and $h_1=h_1(r)$. The odd-parity metric perturbation of the linearized Einstein equations in vacuum,~$\delta G_{\mu \nu}=0$, is then summarized into
\begin{align}
\left(\frac{d^2}{dr^2}+W_1^-\frac{d}{dr}+ W_2^- \right)h=&0.\label{eq:LinearOddeq}
\end{align}
The functions~$W_1^-$ and $W_2^-$  are given by
\begin{widetext}
\begin{align}
\label{eq:P1m}
    W_1^-:=&-\frac{2\left(r-3M\right)r^2}{\sigma^-}\omega^2,\\
    W_2^-:=& W_{20}^-+ W_{22}^- \omega^2+ W_{24}^- \omega^4,\nonumber
\end{align}
with
\begin{align}
    W_{20}^-:=&\frac{\left(\ell+2\right)\left(\ell-1\right)}{r^3 f\sigma^-}\left[ \left(\ell+1\right)\ell r^3-\left\{\left(\ell+1\right)\ell+1\right\}4M r^2+4M^2\left\{\left(\ell+1\right)\ell+4\right\}r-16M^3
  \right],\\
     W_{22}^-:=&-\frac{1}{f\sigma^-}\left[ 2\left\{\left(\ell+2\right)\left(\ell-1\right)-1 \right\}r^2-4M \left\{ \left(\ell+2\right)\left(\ell-1\right)-3\right\}r-16M^2\right],\\
      W_{24}^-:=&\frac{r^4}{f\sigma^-},
\end{align}
where
\begin{align}
    \sigma^-:=& -\left(\ell+2\right)\left(\ell-1\right)f^2r^2 +f r^4\omega^2.
\end{align}
\end{widetext}
We solve Eq.~\eqref{eq:LinearOddeq} perturbatively with small $\omega M$ up to second order. Expand $h$ in $\omega M$ as
\begin{align}
    h=&h^{(0)}+\omega M h^{(1)}+\left(\omega M\right)^2 h^{(2)}+{\cal O}\left(\left(\omega M\right)^3\right).
\end{align}
Then, Eq.~\eqref{eq:LinearOddeq} reduces to
\begin{align}
   {\cal L}_\ell^-\left[h^{(0)} \right]=&0,\label{eq:eqforh0}\\
    {\cal L}_\ell^-\left[h^{(1)}  \right]=&0,\label{eq:eqforh1}\\
     {\cal L}_\ell^-\left[h^{(2)}  \right]=&{\cal S}_\ell^-\left[h^{(0)}\right]\label{eq:eqforh2},
\end{align}
with the derivative operator,
\begin{align}
     {\cal L}_\ell^-:=&\frac{d^2}{dr^2}-\frac{1}{r^2f}\left[\ell\left(\ell+1\right)-\frac{4M}{r}\right],
\end{align}
and the source term,
\begin{align}
\label{eq:Tm}
    {\cal S}_\ell^-:=-\frac{2}{\lambda f^2}\left(\alpha^- r\frac{d}{dr}+ \beta^-\right),
\end{align}
with
\begin{align}
    \alpha^-:=1-\frac{3M}{r},~~ \beta^-:=\frac{\lambda-4}{2}+\frac{6M}{r}.
\end{align}
Here, $\lambda=(\ell+2)(\ell-1)$. In the following, we solve Eqs.~\eqref{eq:eqforh0}--\eqref{eq:eqforh2} order by order.

\subsubsection{Zeroth and first order}
The general solution of Eq.~\eqref{eq:eqforh0} is written in the form,
\begin{align}
    h^{(0)}=&\mathbb{B}_{\ell m}^{(0)} h_\ell^T+\mathbb{I}_{\ell m}^{-(0)} h_\ell^R,\label{eq:generalh00}
\end{align}
with
\begin{align}
     h_\ell^T(r):=&\left(\frac{r}{M}\right)^{\ell+1}\!~_2F_1\left(-\ell+1,-\ell-2;-2\ell;2M/r\right),\nonumber\\
     h_\ell^R(r):=&\left(\frac{M}{r}\right)^\ell \!~_2F_1\left(\ell-1,\ell+2;2\ell+2;2M/r\right).
\end{align}
Here,  $\mathbb{B}_{\ell m}^{(0)}$ and $\mathbb{I}_{\ell m}^{-(0)}$ are dimensionless constants.  

Equation~\eqref{eq:eqforh1} is solved as
\begin{align}
    h^{(1)}=&\mathbb{B}_{\ell m}^{(1)} h_\ell^T+\mathbb{I}_{\ell m}^{-(1)} h_\ell^R.\label{eq:generalh01}
\end{align}
Here, $\mathbb{B}_{\ell m}^{(1)}$ and $\mathbb{I}_{\ell m}^{-(1)}$ are dimensionless constants.

\subsubsection{Second order}
In the same manner as the even-parity case, the general solution of Eq.~\eqref{eq:eqforh2} can be written as
\begin{align}
\label{eq:generalh02}
h^{(2)}=&\mathbb{B}_{\ell m}^{(2)} h_\ell^T+\mathbb{I}_{\ell m}^{-(2)} h_\ell^R+ P_\ell^{-},
\end{align}
where
\begin{align}
\label{eq:Podd}
    P_\ell^-:= \mathbb{B}_{\ell m}^{(0)} P_\ell^{-T}+\mathbb{I}_{\ell m}^{-(0)} P_\ell^{-R},
\end{align}
with
\begin{align}
    P_\ell^{-T}:=&  I_\ell^{-TT} h_\ell^R - I_\ell^{-RT} h_\ell^T ,\\
     P_\ell^{-R}:=&  I_\ell^{-TR} h_\ell^R - I_\ell^{-RR}  h_\ell^T , 
\end{align}
and
\begin{align}
      I_\ell^{-ij}\left(r\right):=\int \frac{h_\ell^i {\cal S}_\ell^-\left[h_\ell^j\right]}{{\cal W}^-}dr,~~{\cal W}^-:=-\frac{2\ell+1}{M},
\end{align}
where $i,j$ run $T,R$; $\mathbb{B}_{\ell m}^{(2)}$ and $\mathbb{I}_{\ell m}^{-(2)}$ are dimensionless constants. 

\subsubsection{General expression for $h$}
We now have the expression for $h$:
\begin{align}
\label{eq:h0}
    h=&\left[\mathbb{B}_{\ell m}^{(0)}+\omega M \mathbb{B}_{\ell m}^{(1)}+\left(\omega M\right)^2 \mathbb{B}_{\ell m}^{(2)}\right]h_\ell^T\nonumber\\
    &+\left[\mathbb{I}_{\ell m}^{-(0)}+\omega M \mathbb{I}_{\ell m}^{-(1)}+\left(\omega M\right)^2\mathbb{I}_{\ell m}^{-(2)}\right]h_\ell^R\\
    &+\left(\omega M\right)^2  P_\ell^{-}+{\cal O}\left(\left(\omega M\right)^3\right).\nonumber 
\end{align}
We also introduce a {\it rescaled magnetic-type tidal response function}:
\begin{widetext}
\begin{align}
\label{eq:rescaledmagtidalresponsefunction}
    \mathscr{F}_{\ell}^-\left(\omega\right):=&-\frac{\ell}{2\left(\ell+1\right)}\left[\frac{\mathbb{I}_{\ell m}^{-(0)}}{\mathbb{B}_{\ell m}^{(0)}}+\left(\omega M\right)\frac{\mathbb{B}_{\ell m}^{(0)}\mathbb{I}_{\ell m}^{-(1)}-\mathbb{B}_{\ell m}^{(1)}\mathbb{I}_{\ell m}^{-(0)}}{\left(\mathbb{B}_{\ell m}^{(0)}\right)^2}\right.\\
    &\left.\quad\quad\quad\quad\quad\quad + \left(\omega M\right)^2\frac{\left(\mathbb{B}_{\ell m}^{(0)}\right)^2 \mathbb{I}_{\ell m}^{-(2)}-\mathbb{B}_{\ell m}^{(0)}\mathbb{B}_{\ell m}^{(1)}\mathbb{I}_{\ell m}^{-(1)}-\mathbb{B}_{\ell m}^{(0)}\mathbb{B}_{\ell m}^{(2)}\mathbb{I}_{\ell m}^{-(0)}+\left(\mathbb{B}_{\ell m}^{-(1)}\right)^2\mathbb{I}_{\ell m}^{-(0)}}{\left(\mathbb{B}_{\ell m}^{-(0)}\right)^3}\right],\nonumber
\end{align}
\end{widetext}
which defines magnetic-type TLNs, TDNs, and dTLNs by 
\begin{align}
\kappa_\ell^{-(0)}:=&{\cal C}^{2\ell+1}  \mathscr{F}_{\ell}^-\left(0\right),\\
\nu_\ell^{-(0)}:=&-i{\cal C}^{2\ell+1} M\frac{d}{d\omega}\mathscr{F}_{\ell m}^-\left(\omega\right)\big|_{\omega\to 0},\\
\kappa_\ell^{-(1)}=&{\cal C}^{2\ell+1} M^2\frac{d^2}{d\omega^2}\mathscr{F}_{\ell m}^-\left(\omega\right)\big|_{\omega\to 0}.
\end{align}

\begin{widetext}
\section{Analytical properties of solutions}\label{Appendix:AnalyticalProperties}
In this appendix, we provide some analytical properties of $H^{(0)}$ in Eq.~\eqref{eq:eqforH0}, $H^{(1)}$ in Eq.~\eqref{eq:eqforH1}, and $H^{(2)}$ in Eq.~\eqref{eq:eqforH2}.

\subsection{Asymptotic expansions of $H^{(0)}$ and $H^{(1)}$}
We first note the analytical properties of Gaussian hypergeometric functions~\cite{NIST:DLMF}. The hypergeometric function is defined by
\begin{align}
    ~\!_2F_1\left(a,b;c;z\right):=\frac{\Gamma\left(c\right)}{\Gamma\left(a\right)\Gamma\left(b\right)}\sum_{s=0}^\infty \frac{\Gamma\left(a+s\right)\Gamma\left(b+s\right)}{s!\Gamma\left(c+s\right)}z^s,
\end{align}
on $|z|<1$, where $\Gamma(z)$ is a gamma function. To investigate the asymptotic behaviors of $H_\ell^T$ and  $H_\ell^R$ near the BH horizon, we use the linear transformations~(15.8.7 and 15.8.10 with 15.8.12 in Ref.~\cite{NIST:DLMF}):\footnote{We use $(-z)_n=(-1)^n(z-n+1)_n$ as well.}
\begin{align}
 ~\!_2F_1\left(-\ell+2,-\ell;-2\ell;2M/r\right)=& \frac{\Gamma\left(\ell+3\right)\Gamma\left(\ell+1\right)}{2\Gamma\left(2\ell+1\right)} ~\!_2F_1\left(-\ell+2,-\ell;3;f\right),\\
  ~\!_2F_1\left(\ell+1,\ell+3;2\ell+2;2M/r\right)=&\frac{\Gamma\left(2\ell+2\right)}{\Gamma\left(\ell+3\right)\Gamma\left(\ell+1\right)} f^{-2} \sum_{k=0}^1 \frac{\left(\ell+1\right)_k \left(\ell-1\right)_k}{k! \left(-1\right)_k} f^k-\frac{\Gamma\left(2\ell+2\right)}{\Gamma\left(\ell+1\right)\Gamma\left(\ell-1\right)}\\
  &\times\sum_{k=0}^\infty \frac{\left(\ell+3\right)_k \left(\ell+1\right)_k}{k!\left(k+2\right)!}f^k \left[ \ln f-\psi\left(k+1\right)-\psi\left(k+3\right)+\psi\left(\ell+1+k\right)+\psi\left(\ell+3+k\right)\right].\nonumber
\end{align}
Here, we have introduced the Pochhammer symbol $(z)_n:=\Gamma(z+n)/\Gamma(z)$ and the digamma function~$\psi(z)$. It follows that $H_\ell^T$ vanishes in the limit $r\to 2M$, while $H_\ell^R$ diverges in the same limit due to the presence of $\ln f$. For instance, $H_2^T$ and $H_2^R$ take the following forms when $\ell=2$:
\begin{align}
H_2^T=f\frac{r^2}{M^2},~~H_2^R=-\frac{5}{16}f\left[ \frac{6}{f^2}\left(\frac{r}{M}-3+\frac{4M}{3r}+\frac{2M^2}{3r^2}\right)+\frac{3r^2}{M^2} \ln f\right].\label{eq:homogeneoussols}
\end{align}
The asymptotic expansions of $H^{(0)}$ and $H^{(1)}$ around $r= 2M$ are
\begin{align}
H^{(0)/(1)}=&\mathbb{E}_{\ell m}^{(0)/(1)}\frac{\Gamma\left(\ell+3\right)\Gamma\left(\ell+1\right)}{2 \Gamma\left(2\ell+1\right)}\epsilon\left[1+{\cal O}\left(\epsilon\right) \right]  \label{eq:asympoticH01} \\
&+ \mathbb{I}_{\ell m}^{+(0)/(1)}\frac{\Gamma\left(2\ell+2\right)}{\Gamma\left(\ell+3\right)\Gamma\left(\ell+1\right)}\epsilon^{-1}\left[1-\left(\ell^2-1\right)f +{\cal O}\left(\epsilon \ln \epsilon\right)\right],\nonumber
\end{align}
where $\epsilon=r/2M-1$.

\subsection{Asymptotic expansions of $H^{(2)}$}
We next provide an explicit form of particular solutions with $\ell=2$. We have 
\begin{align}
P_2^{+T}=&-A_2^{+RT}H_2^T+A_2^{+TT} H_2^R\nonumber\\
&+\frac{2}{315f R^2}\left[72+78R+2694 R^2-4480R^3+\left(3221-420\pi^2f^2\right)R^4-820R^5-660R^6\right.\nonumber\\
&\quad\quad\quad\quad\quad\left. +210\ln \left(R-1\right)\left\{-1-4R+16R^2-8R^3+R^4\left(-2-6f^2\ln\left(R-1\right)\right) \right\}  \right.\label{eq:PT2}\\
&\quad\quad\quad\quad\quad\left.+ 72f^2 R^4\left(-12+35\ln \left(R-1\right)\right)\ln R+2520f^2R^2 {\rm polylog}\left(2,1-R\right)\right],\nonumber
\end{align}
and
\begin{align}
P_2^{+R}=&- A_2^{+RR}H_2^T+A_2^{+TR} H_2^R\nonumber \\
&+\frac{1}{25200f R^2}\left[31500-5250\pi^2+\left(141450-21000\pi^2\right) R-\left(357853-51300\pi^2\right)R^2-\left(458094-23400\pi^2\right)R^3\right.\nonumber\\
&\quad\quad\quad\quad\quad\quad+\left(573747-43200\pi^2\right)R^4+99000R^5\nonumber\\
&\quad\quad\quad\quad\quad\quad-\ln\left(R-1\right)\left\{  16050+32700 R-\left(808650+63000\pi^2\right)R^2+\left(1627500-126000\pi^2\right)R^3 \right.\nonumber\\
&\quad\quad\quad\quad\quad\quad\left.-\left(635100-63000\pi^2\right)R^4 -123000R^5-99000R^6 \right\}\nonumber\\
&\quad\quad\quad\quad\quad\quad+\left(\ln \left(R-1\right)\right)^2\left(15750+63000R-283500 R^2+189000R^3\right) \label{eq:PR2}\\
&\quad\quad\quad\quad\quad\quad-\ln R\left\{ 315000R +\left(519750-63000\pi^2\right)R^2-\left(1434900-126000\pi^2\right)R^3+\left(635100-63000\pi^2\right)R^4\right.\nonumber\\
&\left.\quad\quad\quad\quad\quad\quad +123000R^5+99000R^6 \right\}\nonumber\\
&\quad\quad\quad\quad\quad\quad+ \left\{ \frac{70}{107}\left(\ln R\right)^3 +\frac{35}{107}\left(\ln \left(R-1\right)\right)^3+\ln\left(R-1\right)\ln R\left(1-\frac{105}{107}\ln\left(R-1\right)\right)   \right\} \nonumber \\
&\quad\quad\quad\quad\quad\quad\quad\times\left(192600 R^2-385200R^3+192600 R^4\right)\nonumber\\
&\quad\quad\quad\quad\quad\quad\quad +900\left(35+140R-202R^2-436R^3+428R^4-420 f^2R^4\ln f\right){\rm polylog}\left(2,1-R\right)\nonumber\\
&\left.\quad\quad\quad\quad\quad\quad\quad -756000f^2R^4{\rm polylog}\left(3,f\right)+\zeta\left(3\right)\left(756000R^2-1512000R^3+756000R^4\right)\right].\nonumber 
\end{align}
Here, we have defined $R=r/2M$ and $A_2^{+ij}$ are integration constants. One observes that $P_2^{+T}$ and $P_2^{+R}$ have the homogeneous pieces~\eqref{eq:homogeneoussols} controlled by $A_2^{+ij}$. The asymptotic expansions at large distances in the harmonic radial coordinate take the form,
\begin{align}
    P_2^{+T}\left(\bar{r}\right)\big|_{\bar{r}\gg M}=&-\frac{11\bar{r}^4}{42M^4}-\frac{8\bar{r}^3}{63M^3}+\frac{-630A_2^{+RT}+2971-840\pi^2-1284 \ln \left(\bar{r}/2M\right)}{630}\frac{\bar{r}^2}{M^2}+{\cal O}\left(\bar{r}/M,\ln(\bar{r}/M)M^3/\bar{r}^3\right),\label{eq:largePT}\\
    P_2^{+R}\left(\bar{r}\right)\big|_{\bar{r}\gg M}&=\left(-A_2^{+RR}+\frac{6369}{1600}-\frac{179\pi^2}{168}\right)\left(\frac{\bar{r}^2}{M^2}+\frac{4\bar{r}}{M}+3\right)-\frac{M}{2\bar{r}}-\frac{11M^2}{6\bar{r}^2}+{\cal O}\left(M^3/\bar{r}^3,\ln(\bar{r}/M)M^3/\bar{r}^3\right).\label{eq:largePR}
\end{align}
Note that $A_2^{+TT}$ and $A_2^{+TR}$ appear at ${\cal O}(\ln(\bar{r}/M)M^3/\bar{r}^3)$ as expected from the fact that those are coefficients of the homogeneous parts. 
\end{widetext}

\section{Calibration of a tidal response function: odd-parity sector}\label{Appendix:CalibrationOdd}
In the parallel manner to Sec.~\ref{Sec:Calibration}, we propose calibration of a tidal response function for the odd-parity sector. Let us consider the asymptotic expansion of $P_\ell^{-}$ in Eq.~\eqref{eq:Podd} around $r=2M$,
\begin{align}
    &P_\ell^- \sim \mathbb{B}_{\ell m}^{(0)}A_\ell^-\ln f +\mathbb{I}_{\ell m}^{-(0)}\left[B_\ell^-\left(\ln f\right)^2+D_\ell^-\ln f\right]\nonumber\\
    &+C_\ell^- -\frac{2^{2\ell+3}\left(\ell+1\right)}{\ell}\left(\mathbb{B}_{\ell m}^{(0)}\mu_\ell^T+\mathbb{I}_{\ell m}^{-(0)}\mu_\ell^{-R}\right)+{\cal O}\left(f \right),
\end{align}
where $A_\ell^-$, $B_\ell^-$, and $D_\ell^-$ are ${\cal O}(1)$ specified constants; $C_\ell^-$ is an unspecified constant part at subleading order to logarithmic terms in the asymptotic expansion of $\mathbb{B}_{\ell m}^{(0)}I_\ell^{-TT}+\mathbb{I}_{\ell m}^{-(0)} I_\ell^{-TR}$ around $r=2M$; $\mu_\ell^T$ is given by Eq.~\eqref{eq:muellT} while $\mu_\ell^{-R}$ for the lowest $\ell$ values are given by
\begin{align}
    \mu_2^{-R}=&\frac{25}{576}, \nonumber\\
    \mu_3^{-R}=&\frac{197}{10240}, \\
    \mu_4^{-R}=&\frac{79}{12800}. \nonumber
\end{align}
We require that the asymptotic expansion of $P_\ell^-$ in Eq.~\eqref{eq:Podd} around $r=2M$ does not contain any constant terms at subleading order to the logarithmic terms, yielding
\begin{align}
    C_\ell^- =\frac{2^{2\ell+3}\left(\ell+1\right)}{\ell}\left(\mathbb{B}_{\ell m}^{(0)}\mu_\ell^T+\mathbb{I}_{\ell m}^{-(0)}\mu_\ell^{-R}\right).\label{eq:Cminus}
\end{align}
Similar to the even-parity case, this determines the integration constants of $I_\ell^{-TT}$ and $I_\ell^{-TR}$, specifying the coefficient of the $\bar{r}^{-\ell}$ term in the large-distance asymptotic expansions of $P_\ell^{-T}$ and $P_\ell^{-R}$ in Eq.~\eqref{eq:Podd}. We also demand that the asymptotic expansion of $P_\ell^-$ at large distances contains no terms proportional to $\bar{r}^{\ell+1}$, determining the integration constants of $I_\ell^{-RT}$ and $I_\ell^{-RR}$, and thereby determining the functional form of the particular solutions. When $\kappa_\ell^{-(0)}=0$ as in the BH case, the first condition is sufficient for the unique determination of dTLNs; otherwise, the second is also required.

Next, we redefine tidal moments at the high-frequency regime:
\begin{align}
    \bar{\mathbb{B}}_\ell^{(1)}=\mathbb{B}_{\ell m}^{(1)}-\xi_{\ell m}^{-(1)},~~\bar{\mathbb{B}}_\ell^{(2)}=\mathbb{B}_{\ell m}^{(2)}-\xi_{\ell m}^{-(2)}.
\end{align}
Note that $\xi_{\ell m}^{-(1)}$ and  $\xi_{\ell m}^{-(2)}$ are not necessarily infinitesimal. Then, the induced multipole moments remain invariant if and only if TDNs and dTLNs are redefined,
\begin{align}
    \bar{\nu}_\ell^{-(0)}=&\nu_\ell^{-(0)}- i \kappa_\ell^{-(0)}\frac{\xi_{\ell m}^{-(1)}}{\mathbb{B}_{\ell m}^{(0)}},\\
    \bar{\kappa}_\ell^{-(1)}=&\kappa_\ell^{-(1)}+i \nu_\ell^{-(0)}\frac{\xi_{\ell m}^{-(1)}}{\mathbb{B}_{\ell m}^{(0)}}\\
    &-\kappa_\ell^{-(0)}\left(\frac{\mathbb{B}_{\ell m}^{(1)}-\xi_{\ell m}^{-(1)}}{\left(\mathbb{B}_{\ell m}\right)^2}\xi_{\ell m}^{-(1)}-\frac{\xi_{\ell m}^{-(2)}}{\mathbb{B}_{\ell m}^{(0)}}\right).\nonumber
\end{align}
Choosing $\xi_{\ell m}^{-(1)}=\mathbb{B}_{\ell m}^{(1)}$ and $\xi_{\ell m}^{-(2)}=\mathbb{B}_{\ell m}^{(2)}$, the contributions of tidal moments at the non-adiabatic regime are removed. We thus obtain the calibrated tidal response function in the odd-parity sector:
\begin{align}
\hat{\cal \tilde{F}}_\ell^{-}\left(\omega\right):= \kappa_\ell^{-(0)}+i\omega M \tilde{\nu}_\ell^{-(0)}+\left(\omega M\right)^2 \tilde{\kappa}_\ell^{-(1)}+{\cal O}\left(\omega^3\right),\label{eq:CTRFodd}
\end{align}
with
\begin{align}
    \tilde{\nu}_\ell^{-(0)}:=&\nu_\ell^{-(0)}- i \kappa_\ell^{-(0)}\frac{\mathbb{B}_{\ell m}^{(1)}}{\mathbb{B}_{\ell m}^{(0)}},\label{eq:redifinitionofoddTDNs}\\
    \tilde{\kappa}_\ell^{-(1)}:=&\kappa_\ell^{-(1)}+i \nu_\ell^{-(0)}\frac{\mathbb{B}_{\ell m}^{(1)}}{\mathbb{B}_{\ell m}^{(0)}}+\kappa_\ell^{-(0)}\frac{\mathbb{B}_{\ell m}^{(2)}}{\mathbb{B}_{\ell m}^{(0)}}.\label{eq:redifinitionofodddTLNs}
\end{align}

\section{Computation of tidal response functions}\label{Appendix:TidalResponseFunctions}
In this appendix, we provide the technical details of computing the tidal response function in both even- and odd-parity sectors.
\subsection{Even-parity sector}

\subsubsection{Boundary condition at the BH horizon}
We impose a boundary condition around $r=2M$ on Eq.~\eqref{eq:H0}. To do so, we introduce a series expansion around $r=2M$:
\begin{align}
\label{eq:seriesIng}
    H^{\rm In}=\left(\frac{r}{2M}-1\right)^{-1-2i\omega M}\sum_{j=0}^\infty \alpha_j^+ \left(\frac{r}{2M}-1\right)^j,
\end{align}
which satisfies an ingoing-wave boundary condition around the BH horizon. Substituting this into Eq.~\eqref{eq:LinearEveneq}, all the coefficients~$\alpha_{j}^+ $ are completely determined except for $\alpha_0^+ $. Then, the asymptotic expansion of Eq.~\eqref{eq:seriesIng} in $\omega M$ leads to
\begin{align}
    H^{\rm In}=H^{{\rm In},(0)} +\omega M H^{{\rm In},(1)} +\left(\omega M\right)^2H^{{\rm In},(2)} ,
\end{align}
where the asymptotic behavior of each function near the horizon is given by
\begin{align}
    H^{{\rm In},(0)} :=& -i\frac{\ell\left(\ell^3+2\ell^2-\ell-2\right)}{4}\epsilon \alpha_0^+ \nonumber\\
    &-i\frac{\ell \left(\ell^5+3\ell^4-2\ell^3-9\ell^2+\ell+6\right)}{12}\epsilon^2\alpha_0^+ \label{eq:H0In0}\\
    &+{\cal O}\left(\epsilon^3\right),\nonumber\\
     H^{{\rm In},(1)} :=&\frac{\alpha_0^+}{\epsilon}-\left(\ell^2+\ell-1\right)\alpha_0^+ \nonumber\\
     &-\frac{\epsilon}{4}\left[\ell^4+2\ell^3+3\ell^2+2\ell-12\right.\label{eq:H0In1}\\
     &\left.+2\ell\left(\ell^3+2\ell^2-\ell-2 \right)\ln \epsilon\right]\alpha_0^+ +{\cal O}\left(\epsilon^2\right),\nonumber\\
      H^{{\rm In},(2)} :=& -\frac{2i \ln \epsilon}{\epsilon}\alpha_0^+  \nonumber\\
      &+2i \left[2\ell^2+2\ell+1 +\left(\ell^2+\ell-1\right)\ln \epsilon\right]\alpha_0^+  \label{eq:H0In2}\\
      &+{\cal O}\left(\epsilon \right),\nonumber
\end{align}
where $\epsilon:= r/2M-1$. Here, $\ln \epsilon$ comes from $\epsilon^{-1-2i \omega M}$ in Eq.~\eqref{eq:seriesIng}. In what follows, we determine $\hat{\cal F}_{\ell}(\omega)$ in Eq.~\eqref{eq:H0} at each order in $\omega M$.

\subsubsection{Zeroth order}
Equation~\eqref{eq:H0} at zeroth order in $\omega M$ does not contain $P_\ell^{+T/R}$. As outlined in Appendix~\ref{Appendix:AnalyticalProperties}, $H_\ell^{R}$ contains terms proportional to $1/\epsilon$, while Eq.~\eqref{eq:H0In0} shows that an ingoing-wave solution at this order does not. One can eliminate the term of $1/\epsilon$ by setting $\mathbb{I}_{\ell m}^{+(0)}=0$, which is equivalent to $\hat{\cal F}_{\ell}(0)=0$ from Eq.~\eqref{eq:response} and this leads to
\begin{align}
    \kappa_\ell^{(0)}=0.
\end{align}
This is identical to Eq.~\eqref{eq:zeroFp0}. 

Comparing the coefficient in front of the $\epsilon$ term in Eq.~\eqref{eq:asympoticH01} with $\mathbb{I}_{\ell m}^{+(0)}=0$ with that in Eq.~\eqref{eq:H0In0} allows one to express $\alpha_0^+$ in terms of $\mathbb{E}_{\ell m}^{(0)}$. The coefficient~$\alpha_0^+$ is thus identified as 
\begin{align}
    \alpha_0^+= i\frac{2^{\ell+1}  \Gamma\left(\ell+3\right) \Gamma\left(\ell+1\right)}{\ell\left(\ell^3+2\ell^2-\ell-2\right)\Gamma\left(2\ell+1\right)}\mathbb{E}_{\ell m}^{(0)}.\label{eq:alpha0}
\end{align}

\subsubsection{First order}
Similar to the zeroth-order case, Eq.~\eqref{eq:H0} at first order in $\omega M$ does not contain $P_\ell^{+T/R}$. Based on the $1/\epsilon$ term in Eq.~\eqref{eq:H0In1}, we see that the coefficient of the term of $1/\epsilon$ from $H_\ell^R$ provided in Eq.~\eqref{eq:asympoticH01} should be identical to $\alpha_0^+$ in Eq.~\eqref{eq:alpha0}. This identification determines $\mathbb{I}_{\ell m}^{+(1)}$. From Eq.~\eqref{eq:response}, the rescaled tidal response function at $\mathcal{O}(\omega M)$ is simply 
\begin{equation}
    \mathscr{F}_{\ell}^{+(1)}\left(\omega\right) = \frac{\mathbb{I}_{\ell m}^{+(1)}}{2 \mathbb{E}_{\ell m}^{+(0)}},
\end{equation}
when $\mathbb{I}_{\ell m}^{+(0)}=0$. This leads to
\begin{align}
    \nu_\ell^{(0)}=\frac{\left(\ell+2\right)!\ell!\left(\ell-1\right)!\left(\ell-2\right)!}{2\left(2\ell+1\right)!\left(2\ell-1\right)!},
\end{align}
which is identical to Eq.~\eqref{eq:TDNs}. 

Comparing the coefficient of the term at ${\cal O}(\epsilon)$ in Eq.~\eqref{eq:H0In1} with that in the asymptotic expansion of $H^{(1)}$ around $r=2M$, given by Eq.~\eqref{eq:asympoticH01}, $\mathbb{E}_{\ell m}^{(1)}$ is related to $\mathbb{E}_{\ell m}^{(0)}$ with the use of $\alpha_0^+$ in Eq.~\eqref{eq:alpha0}. We thus obtain
\begin{align}
    \mathbb{E}_{\ell m}^{(1)}=4i\frac{\ell\left(\ell+1\right)\left(\psi\left(\ell\right)+\gamma \right)-\ell^2-1}{\ell\left(\ell+1\right)} \mathbb{E}_{\ell m}^{(0)},
\end{align}
where $\psi(\ell)$ and $\gamma (\simeq 0.57721)$ are a digamma function and Euler's constant, respectively.

\subsubsection{Second order}
The asymptotic expansion of $H$ in Eq.~\eqref{eq:H0} around $r=2M$ contains the term proportional to $\ln \epsilon/\epsilon$ at ${\cal O}((\omega M)^2)$. Equating it to the term of $\ln \epsilon/\epsilon$ in Eq.~\eqref{eq:H0In2} determines $\mathbb{I}_{\ell m}^{+(2)}$. From Eq.~\eqref{eq:response}, the rescaled tidal response function at $\mathcal{O}((\omega M)^2)$ becomes
\begin{equation}
    \mathscr{F}_{\ell}^{+(2)}\left(\omega\right) = \frac{\mathbb{E}_{\ell m}^{+(0)} 
 \mathbb{I}_{\ell m}^{+(2)} - \mathbb{E}_{\ell m}^{+(1)} 
 \mathbb{I}_{\ell m}^{+(1)}}{2 \left(\mathbb{E}_{\ell m}^{+(0)} \right)^2},
\end{equation}
when $\mathbb{I}_{\ell m}^{+(0)}=0$.
 We then find
\begin{align}
    \kappa_\ell^{(1)}=-\left(1+\frac{i \mathbb{E}_{\ell m}^{(1)}}{\mathbb{E}_{\ell m}^{(0)}}\right)\nu_\ell^{(0)}-\frac{{\cal C}^{2\ell+1}C_\ell^{+}}{2\mathbb{E}_{\ell m}^{+(0)}},
\end{align}
where the compactness~${\cal C}=1/2$.  The analytic expression for dTLNs is thus
\begin{widetext}
\begin{align}
    \kappa_\ell^{(1)}=\frac{\left(\ell+2\right)!\ell!\left(\ell-1\right)\left(\ell-2\right)!}{2\ell \left(\ell+1\right)\left(2\ell+1\right)!\left(2\ell-1\right)!}\left[ 4\ell\left(\ell+1\right)\left(\gamma+\psi\left(\ell\right)\right)-\ell\left(5\ell+1\right)-4\right]-\frac{C_\ell^{+}}{2^{2\ell+2} \mathbb{E}_{\ell m}^{(0)}},
\end{align}
\end{widetext}
which is identical to Eq.~\eqref{eq:dTLNs}.

\subsection{Odd-parity sector}

\subsubsection{Boundary condition at the BH horizon}
Similar to the even-parity case, we introduce a series expansion around $r=2M$, which describes an ingoing-wave solution:
\begin{align}
\label{eq:seriesIngh0}
    h^{\rm In}=\left(\frac{r}{2M}-1\right)^{-2i\omega M}\sum_{j=0}^\infty \alpha_j^- \left(\frac{r}{2M}-1\right)^j.
\end{align}
Substituting this into Eq.~\eqref{eq:LinearOddeq}, the coefficients~$\alpha_{j(\ge 1)}^-$ are completely determined with unspecified $\alpha_0^-$ left. Then, the asymptotic expansion of Eq.~\eqref{eq:seriesIngh0} in $\omega M$ leads to
\begin{align}
    h^{\rm In}=h^{{\rm In},(0)} +\omega M h^{{\rm In},(1)} +\left(\omega M\right)^2h^{{\rm In},(2)} ,
\end{align}
with
\begin{align}
    h^{{\rm In},(0)} :=& i\frac{\ell^2+\ell-2}{2}\epsilon \alpha_0^- +i\frac{\left(\ell^2+\ell-2\right)^2}{4}\epsilon^2\alpha_0^-+{\cal O}\left(\epsilon^3\right),\label{eq:h0In0}\\
     h^{{\rm In},(1)} :=&\alpha_0^--\left[\ell^2+\ell-4-\left(\ell^2+\ell-2 \right)\ln \epsilon\right]\epsilon \alpha_0^- +{\cal O}\left(\epsilon^2\right),\label{eq:h0In1}\\
      h^{{\rm In},(2)} :=& -2i\alpha_0^- \ln \epsilon\nonumber\\
      &-i \left[4\ell^2+4\ell-10 -\left(2\ell^2+2\ell-8\right)\ln \epsilon \right.\label{eq:h0In2}\\
      &\left.+\left(\ell^2+\ell-2\right)\left(\ln \epsilon\right)^2\right]\epsilon \alpha_0^- +{\cal O}\left(\epsilon^2 \right),\nonumber
\end{align}
where $\epsilon= r/2M-1$.

\subsubsection{Zeroth order}
Equation~\eqref{eq:h0In0} shows that the ingoing-wave solution does not have logarithmic terms. The general solution~$h^{(0)}$ in Eq.~\eqref{eq:generalh00}, on the other hand, contains $\epsilon\ln \epsilon$ near the BH horizon. To remove such a term, we require $\mathbb{I}_{\ell m}^{-(0)}=0$, which leads to  
\begin{align}
    \kappa_\ell^{-(0)}=0.
\end{align}
This recovers the well-known results of Refs.~\cite{Binnington:2009bb,Poisson:2020vap,Poisson:2021yau}. Then, $\alpha_0^-$ is related to $\mathbb{B}_{\ell m}^{(0)}$ via
\begin{align}
    \alpha_0^-=-i\frac{2^\ell\Gamma\left(\ell+3\right)\Gamma\left(\ell-1\right)}{\ell\left(\ell^2+\ell-2\right)\left(2\ell-1\right)\Gamma\left(2\ell-2\right)}\mathbb{B}_{\ell m}^{(0)}.
\end{align}

\subsubsection{First order}
Matching the coefficients of the term of $\epsilon$ from $h_\ell^R$ at ${\cal O}(\epsilon)$ determines $\mathbb{I}_{\ell m}^{-(1)}$, leading to
\begin{align}
\label{eq:magneticTDNs}
    \nu_\ell^{-(0)}=\frac{\left(\ell+2\right)!\ell!\left(\ell-1\right)!\left(\ell-2\right)!}{2\left(2\ell+1\right)!\left(2\ell-1\right)!}.
\end{align}
This is identical to the electric-type TDNs in Eq.~\eqref{eq:TDNs}, being in perfect agreement with Refs.~\cite{Poisson:2020vap,Chia:2020yla,Katagiri:2023yzm,Chakraborty:2023zed}. The coefficient~$\mathbb{B}_{\ell m}^{(1)}$ is determined as
\begin{align}
\label{eq:B1}
    \mathbb{B}_{\ell m}^{(1)}=4i\left(\gamma+\psi\left(\ell\right)+\frac{\ell^3-3\ell-1}{\left(\ell+2\right)\left(\ell+1\right)\ell\left(\ell-1\right)}\right)\mathbb{B}_{\ell m}^{(0)},
\end{align}
where $\psi(\ell)$ and $\gamma (\simeq 0.57721)$ are a digamma function and Euler's constant, respectively.

\subsubsection{Second order}
The asymptotic expansion of $h$ in Eq.~\eqref{eq:h0} around $r=2M$ contains the term proportional to $\ln \epsilon$. Equating this with the term of $\ln \epsilon $ in Eq.~\eqref{eq:h0In2} determines $\mathbb{I}_{\ell m}^{-(2)}$, thereby obtaining the magnetic-type dTLNs,\begin{widetext}
\begin{align}
    \kappa_\ell^{-(1)}=\frac{1}{\ell\left(\ell+1\right)}\left[4\ell\left(\ell+1\right) \left(\gamma+\psi\left(\ell\right)\right)-\frac{2\ell\left(\ell^3-\ell+4\right)+4}{\left(\ell+2\right)\left(\ell-1\right)} \right]\nu_\ell^{-(0)}+\frac{\ell}{2^{2\ell+2}\left(\ell+1\right)\mathbb{B}_{\ell m}^{(0)}}C_\ell^{-}.
\end{align}
\end{widetext}

\subsubsection{Zero magnetic-type dTLNs within the calibration}
Within the calibration proposed in Appendix~\ref{Appendix:CalibrationOdd}, we show vanishing magnetic-type dTLNs. The condition~\eqref{eq:Cminus} leads to
\begin{align}
    C_\ell^{-TT}=\frac{2^{2\ell+3}\left(\ell+1\right)}{\ell}\mathbb{B}_{\ell m}^{(0)}\nu_\ell^{-(0)},
\end{align}
where $\mu_\ell^T=\nu_\ell^{-(0)}$ was used. We then obtain
\begin{align}
\label{eq:magneticdTLNs}
    \kappa_\ell^{-(1)}=-i\frac{\mathbb{B}_{\ell m}^{(1)}}{\mathbb{B}_{\ell m}^{(0)}}\nu_\ell^{-(0)}.
\end{align}

Next, since $\kappa_\ell^{-(1)}=0$, TDNs remain invariant under the redefinition~\eqref{eq:redifinitionofoddTDNs} while the dTLNs in Eq.~\eqref{eq:magneticdTLNs} shift by Eq.~\eqref{eq:redifinitionofodddTLNs}, yielding
\begin{align}
\tilde{\kappa}_\ell^{-(1)}=0.
\end{align}

\bibliography{apssamp}

\end{document}